\documentclass{article}

\usepackage[cmex10]{amsmath}

\usepackage{graphicx}
\usepackage{dcolumn}
\usepackage[utf8]{inputenc}
\usepackage{hyperref}
\usepackage[caption=false]{subfig}
\usepackage{amsmath, amsthm, bm}
\usepackage{amsfonts, amssymb}
\usepackage{newfloat}
\usepackage{geometry}

\newcommand{\G}{\mathcal{G}}
\newcommand{\R}{\mathbb{R}}
\newcommand{\A}{\mathcal{A}}
\newcommand{\Sc}{\mathcal{S}}
\newcommand{\E}{\mathcal{E}}

\newcommand{\M}{\mathcal{M}}
\newcommand{\T}{\mathcal{T}}

\newcommand{\X}{\mathcal{X}}

\DeclareGraphicsExtensions{.jpg,.mps,.pdf,.png} 
\graphicspath{{tex/images/}, {images/}}

\begin{document}

\title{Duality between Temporal Networks and Signals: Extraction of the 
Temporal Network Structures}

\author{Ronan~Hamon, Pierre~Borgnat, Patrick~Flandrin and~C\'eline~Robardet}%

\date{\today}

\maketitle

\begin{abstract}
We develop a framework to track the structure of temporal networks with a signal 
processing approach. The method is based on the duality between networks and 
signals using a multidimensional scaling technique. This enables a study of the 
network structure using frequency patterns of the corresponding signals. An 
extension is proposed for temporal networks, thereby enabling a tracking of the 
network structure over time. A method to automatically extract the most 
significant frequency patterns and their activation coefficients over time is 
then introduced, using nonnegative matrix factorization of the temporal spectra. 
The framework, inspired by audio decomposition, allows transforming back these 
frequency patterns into networks, to highlight the evolution of the underlying 
structure of the network over time. The effectiveness of the method is first 
evidenced on a toy example, prior being used to study a temporal network of 
face-to-face contacts. The extraction of sub-networks highlights significant 
structures decomposed on time intervals.
\end{abstract}

\section{Introduction}

Many complex systems, whether physical, biological or social, can be naturally 
represented as networks, i.e., a set of relationships between entities. Network 
science \cite{Newman2010} has been widely developed to study such objects, 
generally supported by a graph structure, by providing powerful tools, such as 
the detection of communities \cite{Fortunato2010}, in order to understand the 
underlying properties of these systems. Recently, connections between signal 
processing and network theory have emerged: The field of signal processing over 
networks \cite{Shuman2013, Sandryhaila2014} has been introduced with the 
objective of transposing concepts developed in classical signal processing, such 
as Fourier transform or wavelets, in the graph domain. These works have led to 
significant results, among them filtering of signals defined over a network 
\cite{Shuman2013} or multiscale community mining using graph wavelets 
\cite{Tremblay2014}. These approaches benefit from the natural representation of 
networks by graphs, enabling the use of the comprehensive mathematical 
understanding of graph theory. Another approach has been also considered, 
defining a duality between graphs and signals: methods have been developed to 
transform graphs into signals and conversely, in order to take advantage of both 
signal processing and graph theory. Hence, mapping a graph into time series has 
been performed using random walks \cite{Weng2014, Campanharo2011, Girault2014} 
or deterministic methods based on classical multidimensional scaling 
\cite{Haraguchi2009, Shimada2012}. This latter approach has been the topic of 
several extensions in \cite{Hamon2015b}, in order to build a comprehensive 
framework to link frequency patterns of the so-obtained signals with network 
structures.

Studies mainly focused on the analysis of static networks, potentially 
aggregated over time, but without time evolution. However, the considered 
systems are most of the time not frozen: vertices and edges appear and disappear 
over the course of time. Aggregating the networks over a time interval gives 
insight of the underlying mechanisms, but often does not provide the actual 
dynamic sequence of appearance and disappearance of edges: two edges, active one 
after the other, will be considered simultaneous in the temporally-aggregated 
network. Given the importance of knowing such dynamics, for instance in topics 
such as epidemic spread or communication networks, and thanks to the recent 
availability of many data sets, a temporal network theory has recently appeared 
\cite{Holme2012, Casteigts2012}, enabling a deeper understanding of underlying 
mechanisms. Several studies proposed an extension of the methods developed for 
static networks to the temporal case: we can cite for instance works on network 
modeling \cite{Braha2009, Clementi2009, Grindrod2011, Xu2013a}, detection of 
communities \cite{Mucha2010, Xu2011a, Gauvin2014, Gauvin2015, Chen2013}, 
detection of temporal motifs \cite{Kovanen2011}, visualization \cite{Xu2013b}, 
or more generally data mining of time-evolving sequences \cite{Yi2000}.

We propose in this article a new approach based on the duality between networks 
and signals to study temporal networks, introduced in \cite{Hamon2015b}. The 
objective here is to follow the global structure of the network over time. One 
first contribution consists of an extension of the existing work to the temporal 
case, which is naturally performed by considering each time instant as a static 
network. This approach enables us to visually track temporal networks by 
following the frequency patterns associated to specific structures. The second 
contribution consists of using nonnegative matrix factorization (NMF) 
\cite{Lee1999} to automatically extract and track the significant frequency 
patterns over time. Details about the reconstruction how components can be 
transformed back into networks are also provided.

Preliminary versions of this work have been presented in different places: the 
principle of extension of the method to the temporal case is introduced in 
\cite{Hamon2013a}, as well as the visual tracking of frequency patterns 
representing structures. In \cite{Hamon2013b, Hamon2013c}, this approach has 
been used to study a bike sharing system in Lyon, which is not mentioned in this 
article. Finally, the idea of the decomposition using NMF has been suggested in 
\cite{Hamon2014a}. This paper extends however those previous works by detailing 
a comprehensive framework and setting out consistent arguments for the validity 
of the method in the context of analysis of temporal networks.

The paper is organized as follows. Section~\ref{sec:preliminaries} recalls the 
duality between static networks and signals, as introduced in \cite{Shimada2012} 
and extended in \cite{Hamon2015b}. Section~\ref{sec:extension} gives an 
extension of the transformation to the temporal case, and introduces a toy 
temporal network highlighting specific network structures. 
Section~\ref{sec:spectral} shows how spectral analysis is used to track the 
structure of the temporal network over time, while 
Section~\ref{sec:decomposition} describes a decomposition of a temporal network 
into sub-networks by applying nonnegative matrix factorization to the matrix of 
spectra over time. For both sections, an illustration with the toy temporal 
network is provided. Finally, Section~\ref{sec:primary} applies the developed 
framework to a real-world temporal network from the SocioPatterns project, 
describing face-to-face contact between children in a primary school.

\paragraph*{\textbf{Notations}}

Matrices are denoted by boldface capital letters, their columns by boldface 
lowercase letters, and their elements by lowercase letters: for $\bm{M} \in 
\R^{A\times B}$, $\bm{M} = [\bm{m}_1, \hdots, \bm{m}_B] = (m_{ab})_{a \in \{1, 
\hdots, A\}, b \in \{1, \hdots, B\}} $. Tensors are represented by calligraphic 
capital letters: for $\T \in \R^{A\times B \times T}$, $\T =[\bm{M}^{(t)}]_{t 
\in \{1, \hdots, T\}}$. Operators are represented by calligraphic boldface 
capital letters: $\bm{\mathcal{F}}$.


\section{Duality between static networks and signals}
\label{sec:preliminaries}

\subsection{Transformation from static networks to signals}

Duality between networks and signals has been introduced to analyze networks 
structures using signal processing tools. Shimada et al.~\cite{Shimada2012} 
proposed a method to transform static networks into a collection of signals 
using multidimensional scaling. We proposed in \cite{Hamon2015b} a comprehensive 
framework to transform graphs into a collection of signals, based on this work. 
We recall in this section the main principles of this framework, which will be 
used in the following to study temporal networks. We consider in the following 
networks described by unweighted and undirected graph with $N$ vertices.

In \cite{Shimada2012}, Classical MultiDimensional Scaling (CMDS) \cite{Borg2005} 
is used to transform a graph into signals by projecting the $N$ vertices of the 
graph in a Euclidean space, such that distances between these points correspond 
to relations in the graph. The graph is fully described by its adjacency matrix 
$\bm{A} \in \R^{N \times N}$, whose elements $a_{ij}$ are equal to $1$ if 
vertices $i$ and $j$ are linked, and $0$ otherwise. The transformation consists 
in applying CMDS to a matrix encoding distance between vertices of a graph, 
noted 
$\mathbf{\Delta} = (\delta_{ij})_{i,j=1,..,N}$, and defined for two vertices 
$i,j$ of the network by:
\begin{align}
  \delta_{ij} = \left\{
  \begin{array}{l l}
    0 & \quad \text{if $i=j$}\\
    1& \quad \text{if $a_{ij} = 1$ and $i\neq j$}\\
     w > 1 & \quad \text{if $a_{ij} = 0$ and $i\neq j$}\\
  \end{array} \right.
\end{align}
As discussed in \cite{Hamon2015b}, we choose $w=1 + \frac{1}{N}$. This 
definition focuses on the presence (denoted by a distance equal to $1$) or the 
absence (denoted by a distance equal to $w$) of an edge between two vertices. 
Hence, the distance of two vertices in the graph, often defined as the length of 
the shortest path between the two vertices, has no direct impact on the 
matrix~$\bm{\Delta}$: two pairs of unlinked vertices will have a distance equal 
to $w$, whether they are close or not in the graph. Applying CMDS on the 
distance matrix~$\bm{\Delta}$ leads to a collection of points, corresponding to 
the vertices, in a Euclidean space $\R^{N-1}$. The considered signals, or 
components, correspond to the coordinates, for each dimension, of the points. 
The collection of signals is denoted by $\bm{X} \in \R^{N \times C}$, where $C$ 
is the total number of components, and thus is equal to $N-1$ for the full 
representation. The columns $\bm{x}_{c}$ represent the $c$-th signal, with $c 
\in \{1, \hdots C\}$, and are indexed by the vertices. The transformation is 
denoted by $\bm{\T}$: $\bm{\T}[\bm{A}] = \bm{X}$.

\subsection{Inverse Transformation}

Transforming back signals to a graph has to take into account the nature of the 
signals, as they are a representation of a specific network. The inverse 
transformation must hence preserve the original topology of the underlying 
graph. By construction of the collection of signals $\bm{X}$, the perfect 
retrieval of the underlying graph is easily reachable, by considering the 
distances between each pair of point: As built using CMDS, these distances 
represent the distance matrix $\bm{\Delta}$, and the adjacency matrix of the 
graph is directly obtained. However, when $\bm{X}$ is degraded or modified, e.g. 
by taking $C <N-1$, the distances are no longer directly the ones computed 
between vertices, even if they stay in the neighborhood of these distances. We 
proposed in \cite{Hamon2015b} to take into account the energy of components to 
improve the reconstruction, as well as prior information about the original 
graph. If the distance between two vertices $i$ and $j$ in a high-energy 
component is high, it means that the two vertices are likely to be distant in 
the graph. Conversely, if the distance in a high-energy component is low, then 
the two vertices are likely to be connected in the graph. 

Let $\tilde{\bm{X}}$ be a degraded collection of signals. The energies are 
normalized according to the energies of the original components, by multiplying 
the components $\tilde{\bm{x}}_{c}$ by the normalization factor $N_c$:
\begin{align} 
N_c = \sqrt{\frac{\sum_{n=1}^N x_{nc}^2}{\sum_{n=1}^N \tilde{x}_{nc}^2}}
\end{align}
Then the distances are computed by using the energies as follows:
\begin{align}
\label{eq:weighted_distances} 
d(\bm{X})_{ij} = \sqrt{\sum_{c=1}^C {(u_c)^\alpha(x_{ic} - x_{jc})^2}}
\end{align}
with $\alpha \geq 0$,  where $u_c$ is the energy of component $c$, computed as 
$u_c = \sum_{i=1}^n {x_{ic}^2}$, and normalized such that $\|\bm{u}\|_\alpha = 
\sum_{c=1}^C (u_c)^\alpha = C$, with $\bm{u}$ the vector of energies for all 
components. The parameter $\alpha$ controls the importance of the weighting: if 
$\alpha$ is high, the high-energy components have a higher importance in the 
computation of distances compared to the low-energy components. Conversely, if 
$\alpha$ is small, the importance of high-energy components is diminished. In 
particular, $\alpha=0$ gives the standard reconstruction. The distances are then 
thresholded, by retaining the edges corresponding to the smallest distances. The 
threshold is chosen in order to recover the same amount of edges than in the 
original network. This inverse transformation is denoted $\bm{\T^{-1}}$: 
$\bm{\T^{-1}}[X] = \bm{A}^{(r)}$, where $\bm{A}^{(r)}$ is the adjacency matrix 
of the reconstructed graph.

\subsection{Indexation of signals}

As the signals are indexed by the vertices, the order in which we consider them 
in the transformation is essential to study some aspects of the signals, 
especially when using spectral analysis of the signals. Ordering randomly the 
vertices does not change the value assigned to each vertex, but would lead to 
abrupt variations in the representation of signals: Specific frequency 
properties, clearly observable in signals, will no longer be visible. 
Unfortunately, the suitable ordering is usually not available, especially when 
dealing with real-world graphs. To address this issue, we proposed in 
\cite{Hamon2015a} to find a vertex ordering that reflects the topology of the 
underlying graph, based on the following assumption: if two vertices are close 
in the graph (by considering for instance the length of the shortest path 
between them), they have to be also close in the ordering. Details of the 
algorithm and results about the consistency between the obtained vertex ordering 
and the topology of the graph are covered in \cite{Hamon2015a}.

\subsection{Spectral analysis of signals}
\label{subsec:spectral_analysis}

Spectral analysis is performed using standard signal processing methods: Let a 
collection $\bm{X}$ of $C$ signals indexed by $N$ vertices. The spectra $\bm{S} 
\in \R^{C \times F}$ give the complex Fourier coefficients, whose elements are 
obtained by applying the Fourier transform on each of the $C$ components of 
$\bm{X}$:
\begin{align}
	\bm{s}_c = \bm{\mathcal{F}}[\bm{x}_c]
\end{align}
estimated, for positive frequencies, on $F = \frac{N}{2} + 1$ bins, 
$\bm{\mathcal{F}}$ being the Fourier transform and $c \in \{1, \hdots, C \}$. 

From the spectrum $\bm{S}$, the following features are obtained for each 
frequency of each component:
\begin{itemize}
	\item the magnitudes $\bm{M}$, which read as $m_{cf} = |s_{cf}|$
	\item the energies $\bm{E}$, which read as $e_{cf} = |s_{cf}|^2$
	\item the phases $\bm{\Phi}$, which read as $\phi_{cf} = \arg(s_{cf})$
\end{itemize}

The matrices $\bm{M}$ and $\bm{E}$ are studied as a frequency-component map, 
exhibiting patterns in direct relation with the topology of the underlying 
graph. The phases of signals $\bm{\Phi}$ are used in the inverse Fourier 
transformation, when the collection of signals has to be retrieved from $\bm{M}$ 
or $\bm{E}$.

\subsection{Illustrations}
\label{subsec:illustrations}

\begin{figure}

	\subfloat[\label{subfig:gs_rl-100-4}$4$-ring lattice]{
		\includegraphics[width=0.31\columnwidth]{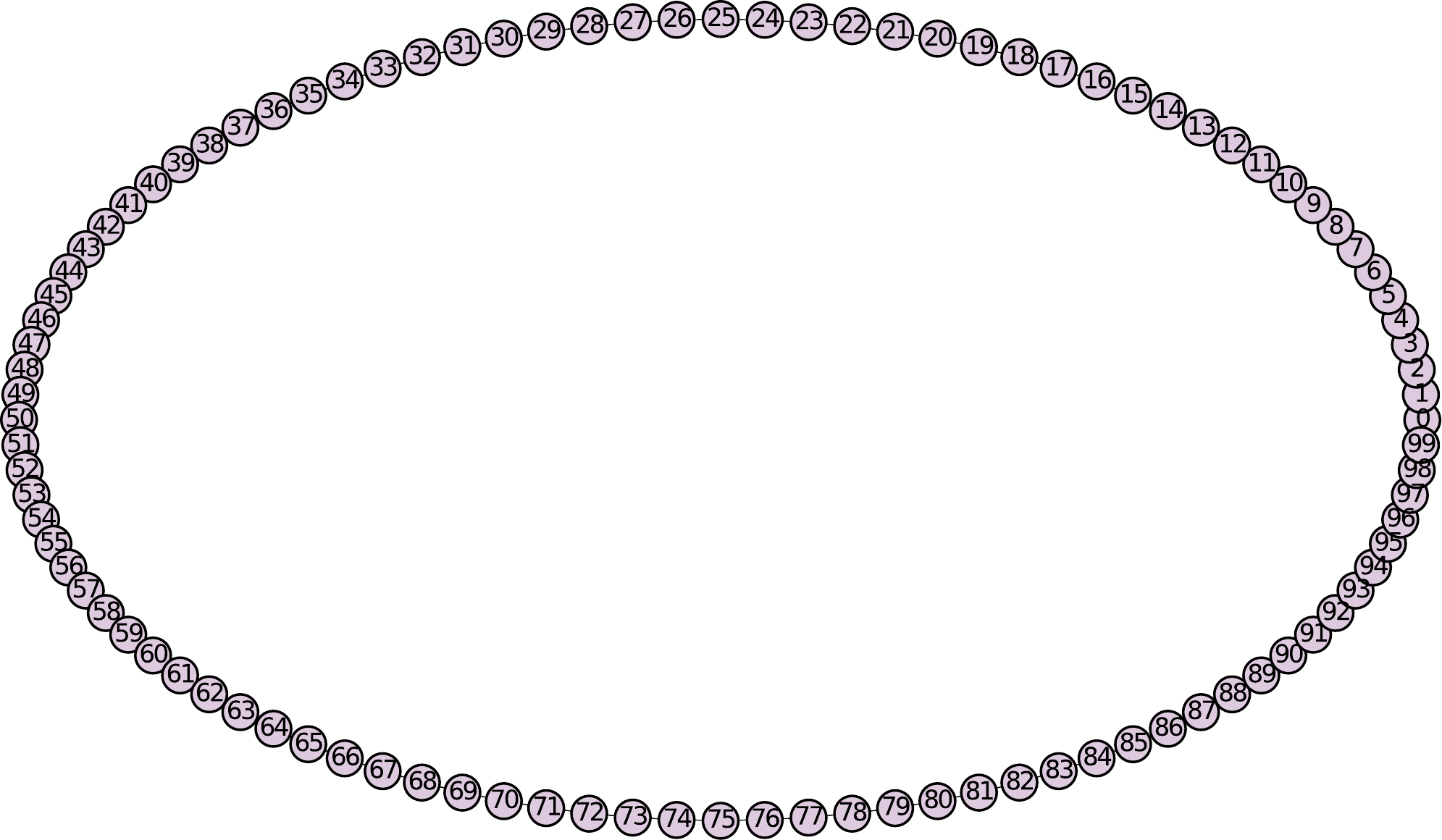}
		\includegraphics[width=0.31\columnwidth]{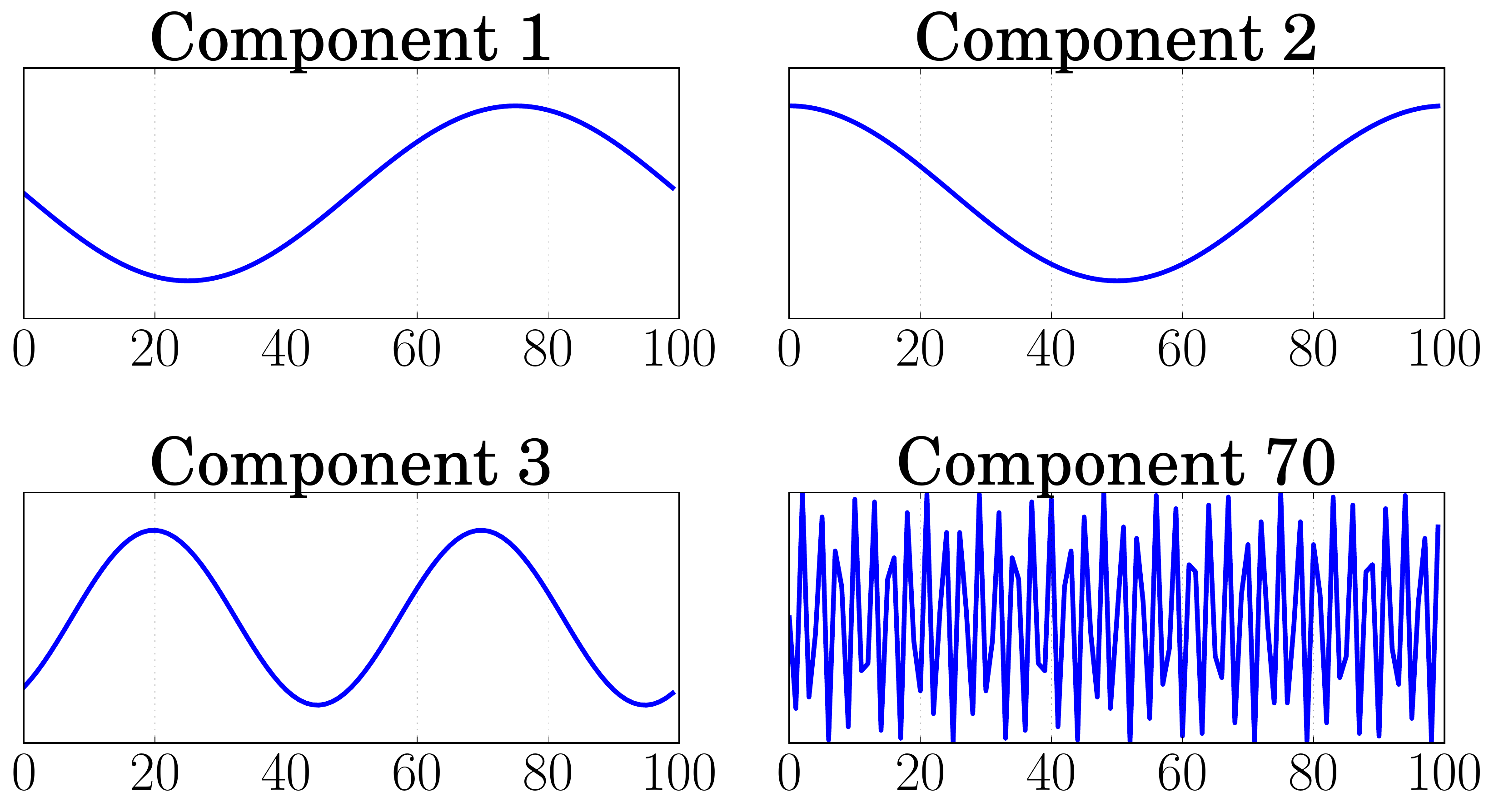}
		\includegraphics[width=0.31\columnwidth]{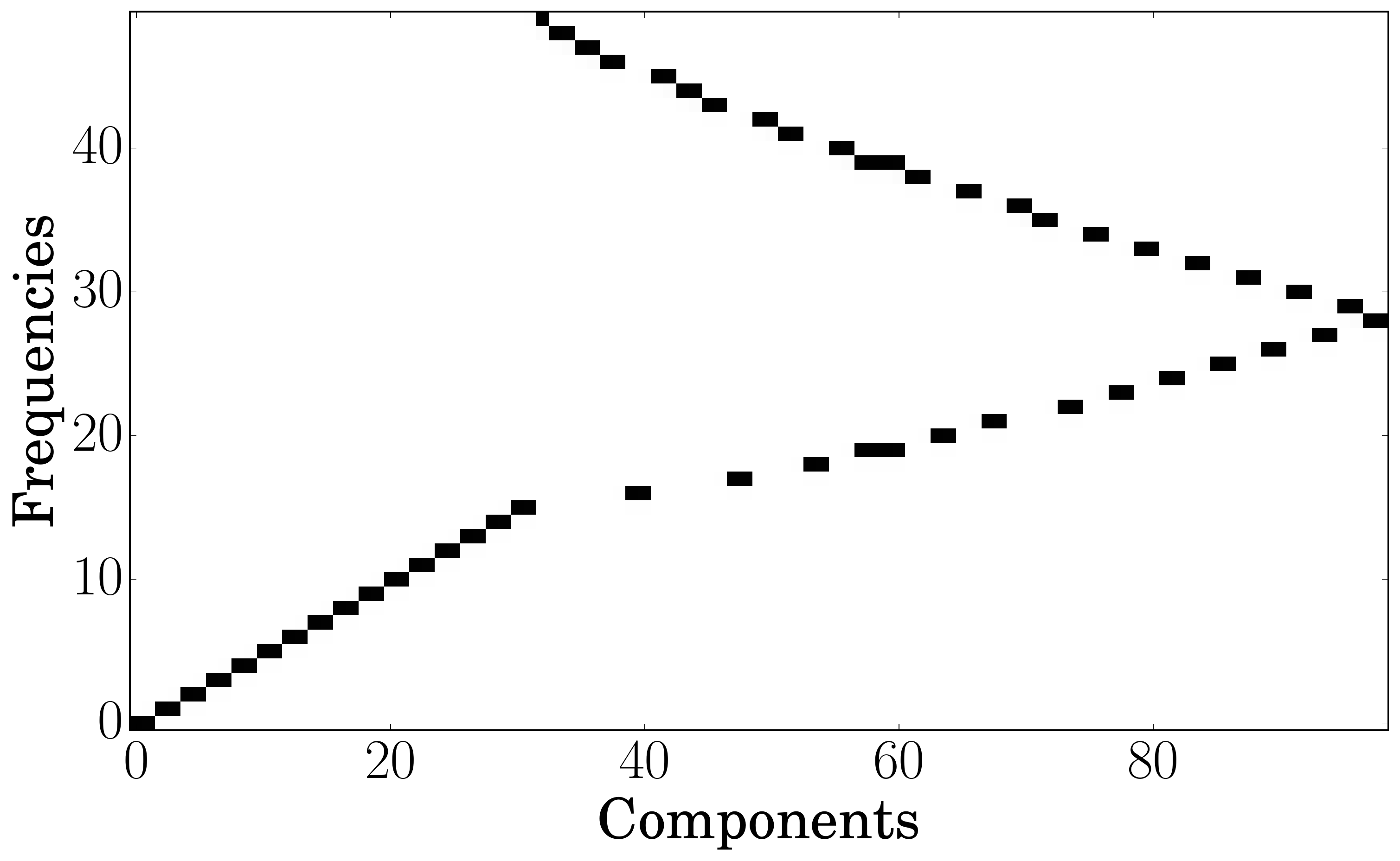}
	}
	
	\subfloat[\label{subfig:gs_nrl-100-4-1}Noised $4$-ring lattice with 
$p=0.1$]{
		\includegraphics[width=0.31\columnwidth]{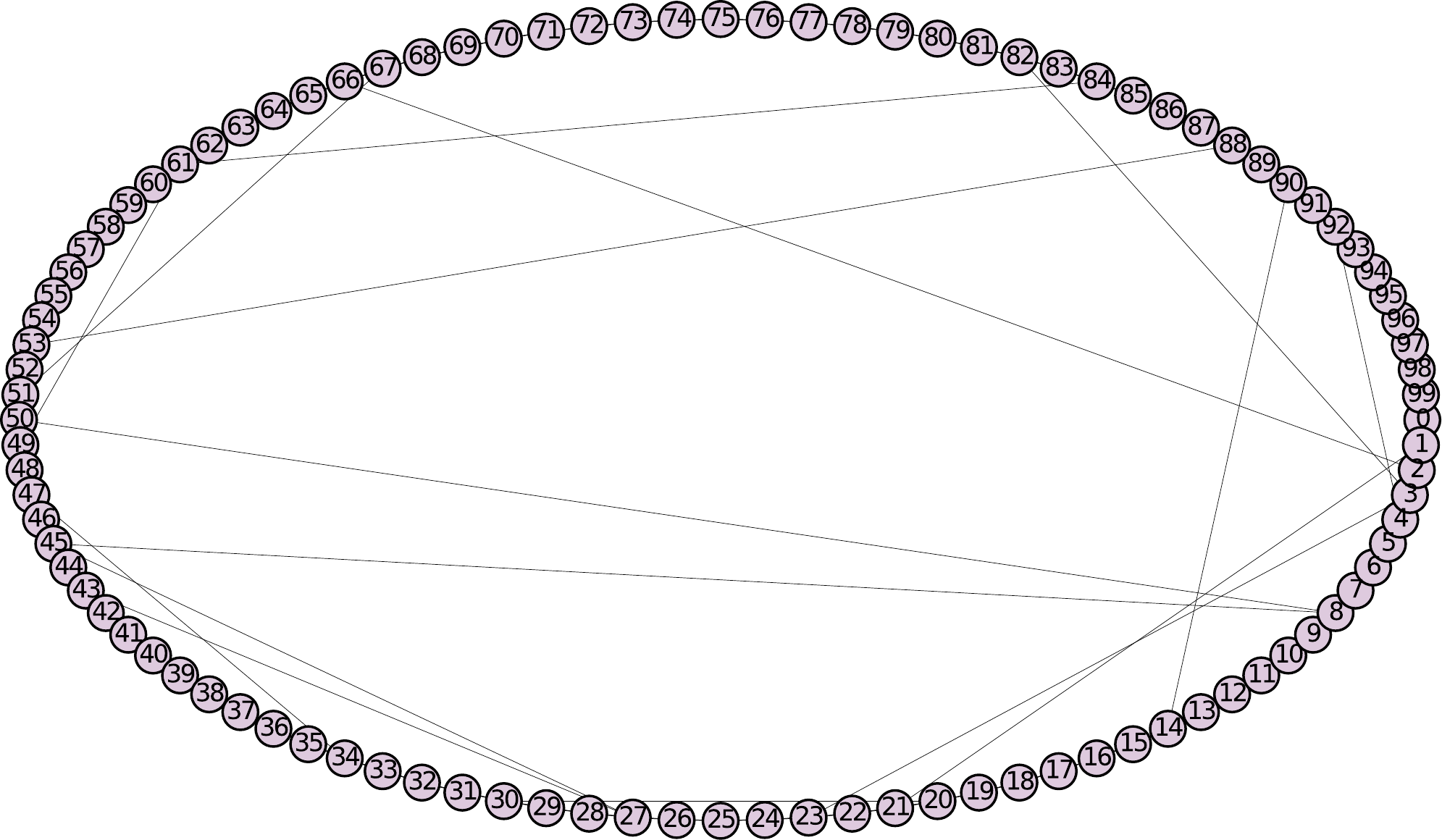}
		\includegraphics[width=0.31\columnwidth]{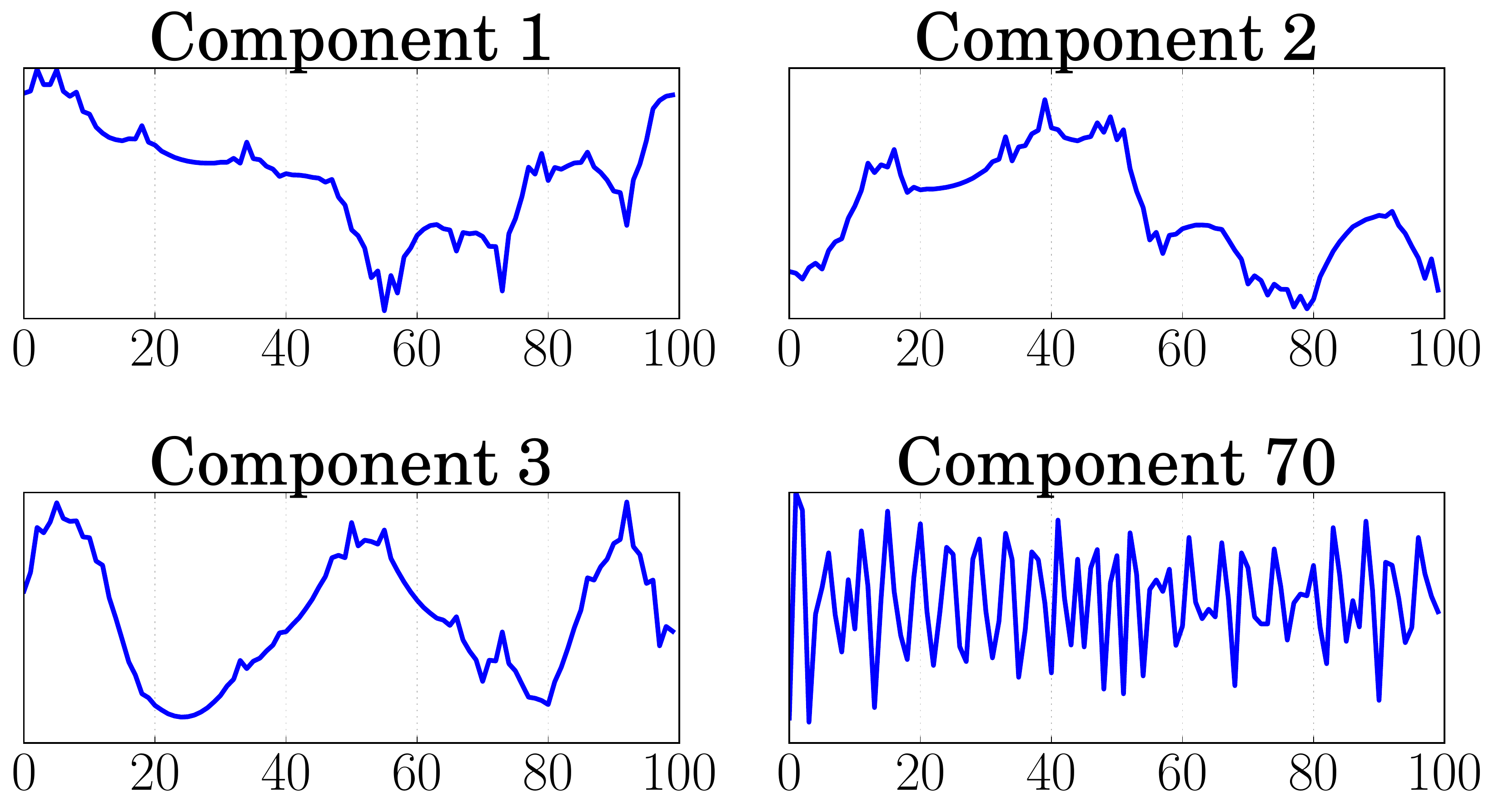}
		\includegraphics[width=0.31\columnwidth]{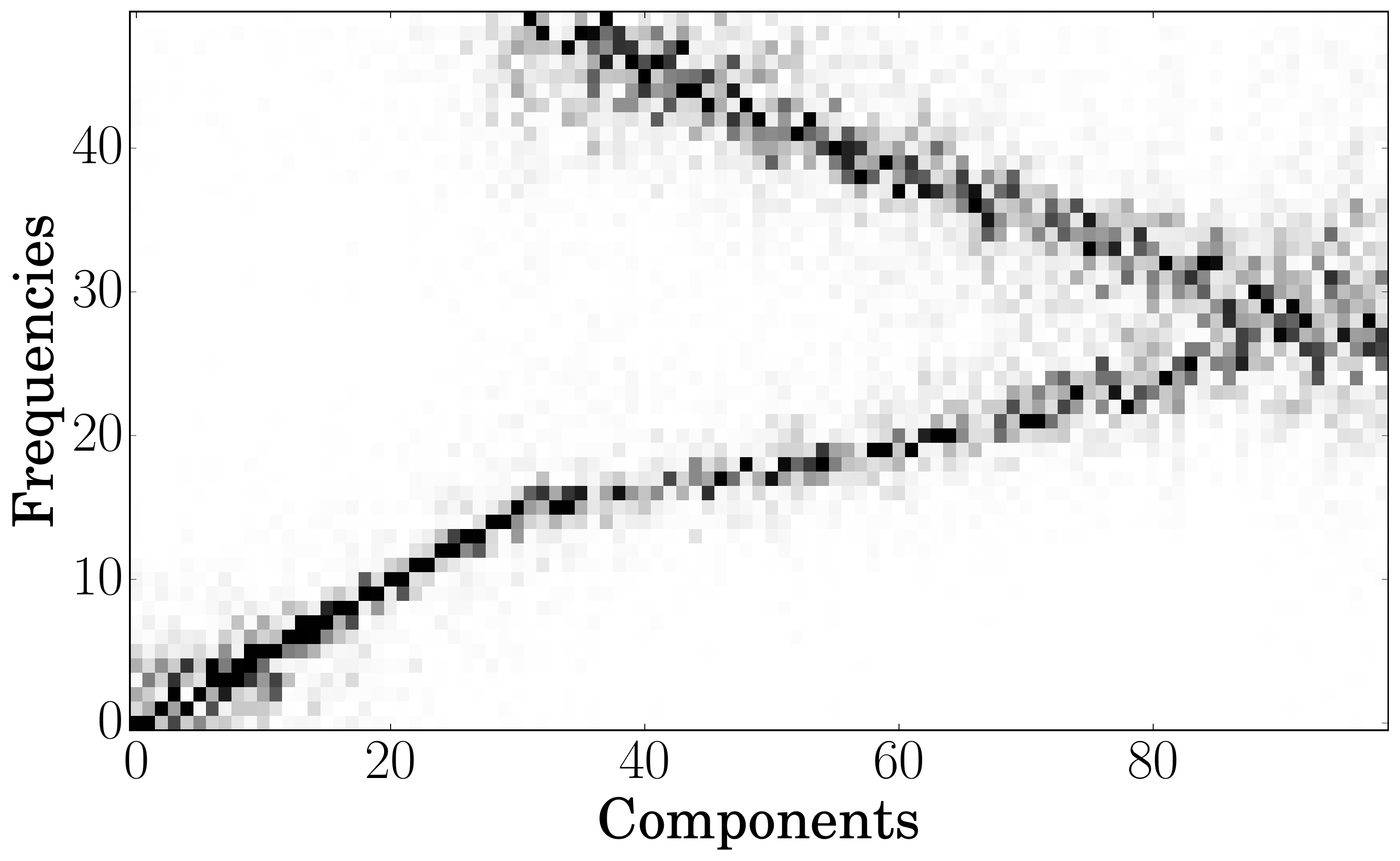}
	}
	
	\subfloat[\label{subfig:gs_sbm-100-3}Stochastic block model with $3$ 
communities, $p_w=0.8$ and $p_b=0.05$]{
		\includegraphics[width=0.31\columnwidth]{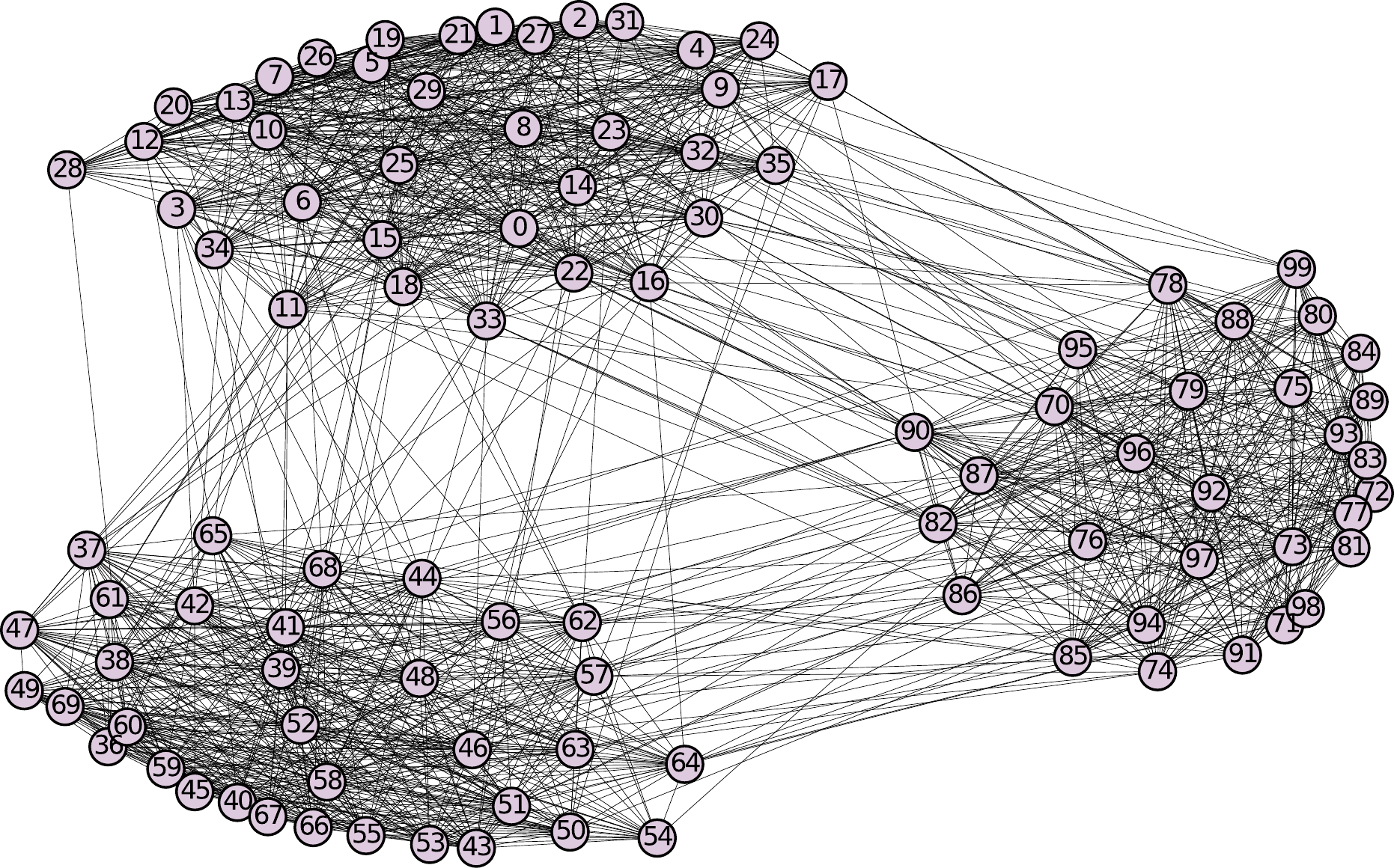}
		\includegraphics[width=0.31\columnwidth]{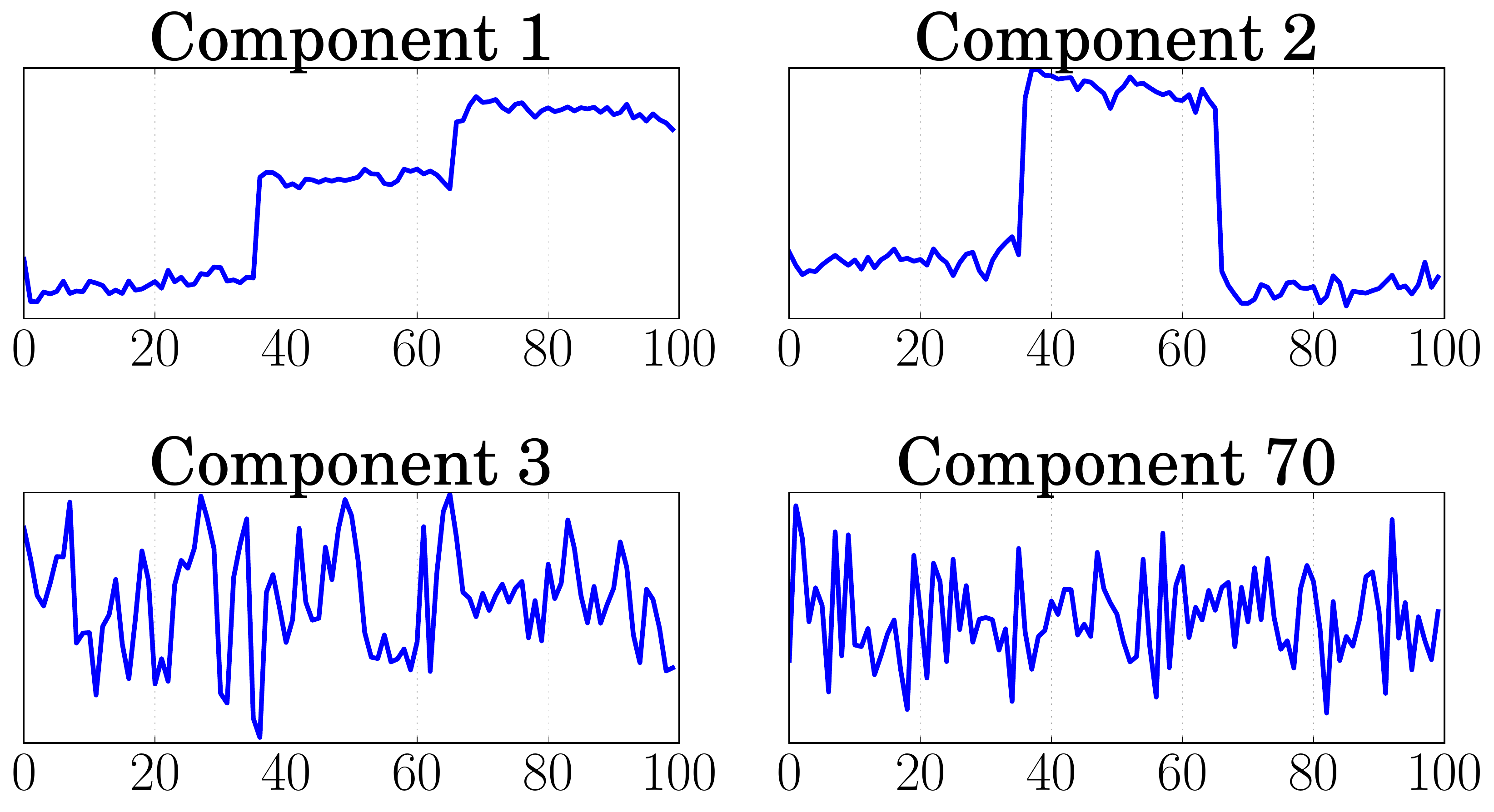}
		\includegraphics[width=0.31\columnwidth]{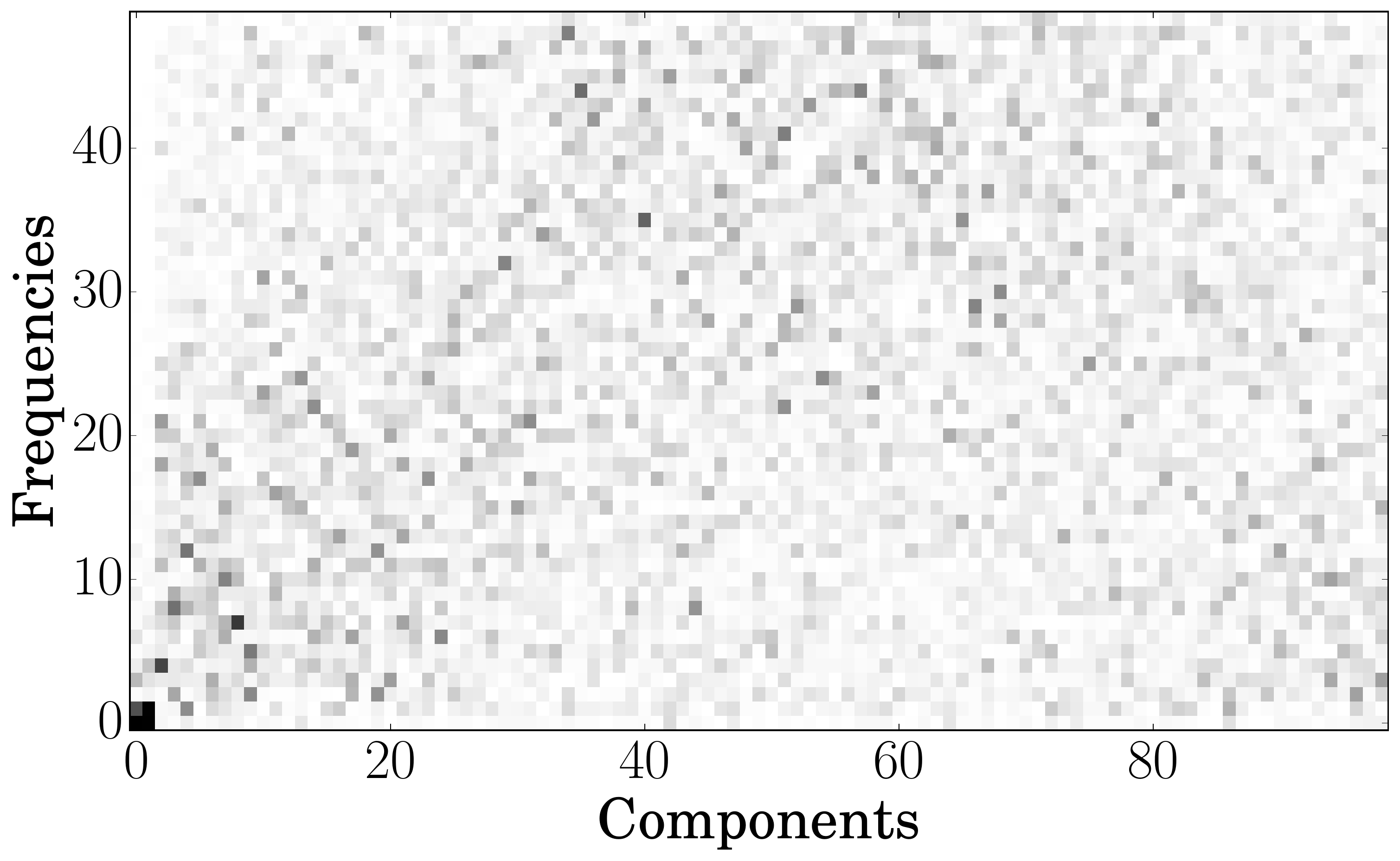}
	}
	
	\subfloat[\label{subfig:gs_rl_sbm-100-4-3}$4$-ring lattice with $3$ 
communities]{
		\includegraphics[width=0.31\columnwidth]{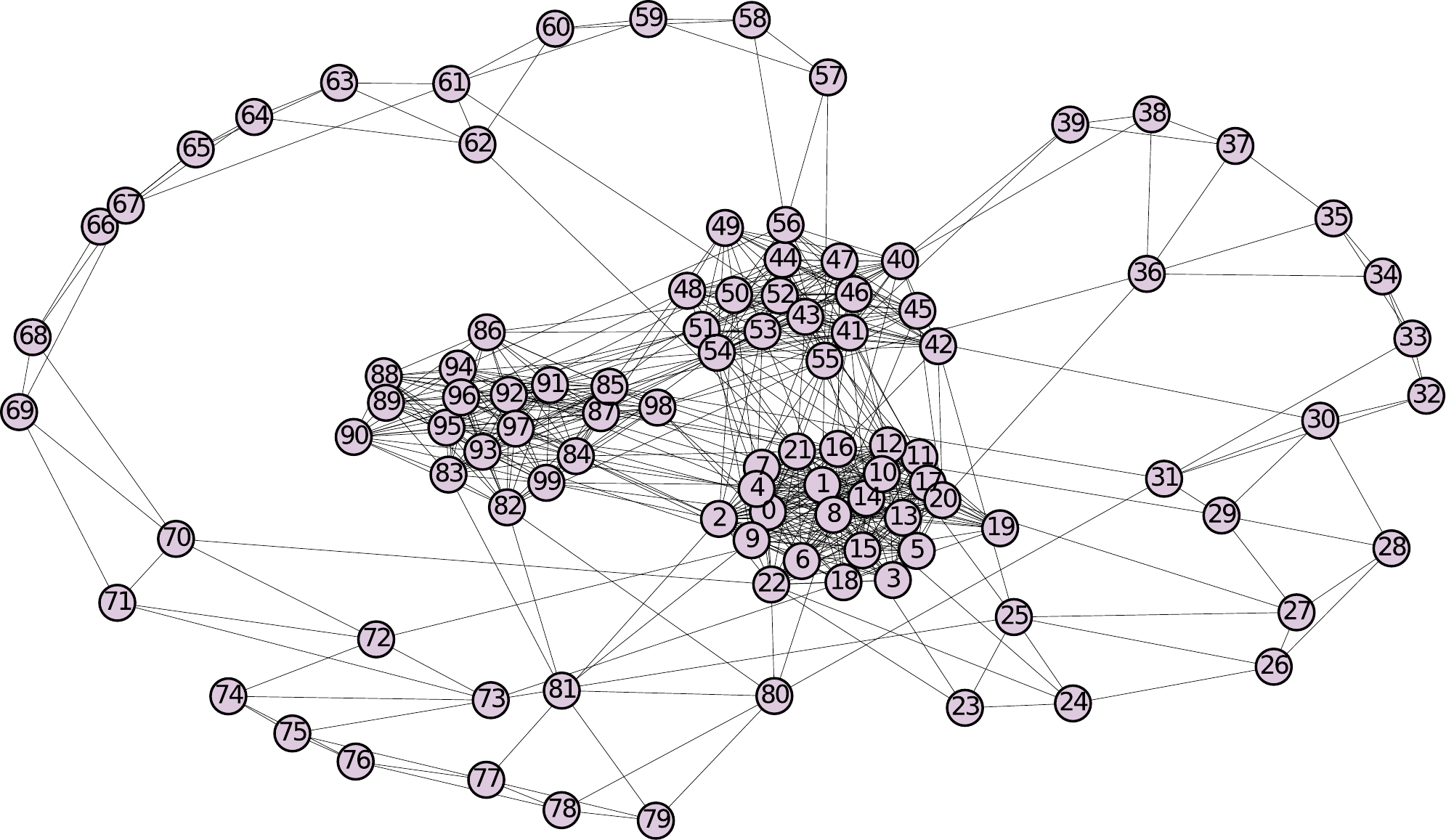}
		\includegraphics[width=0.31\columnwidth]{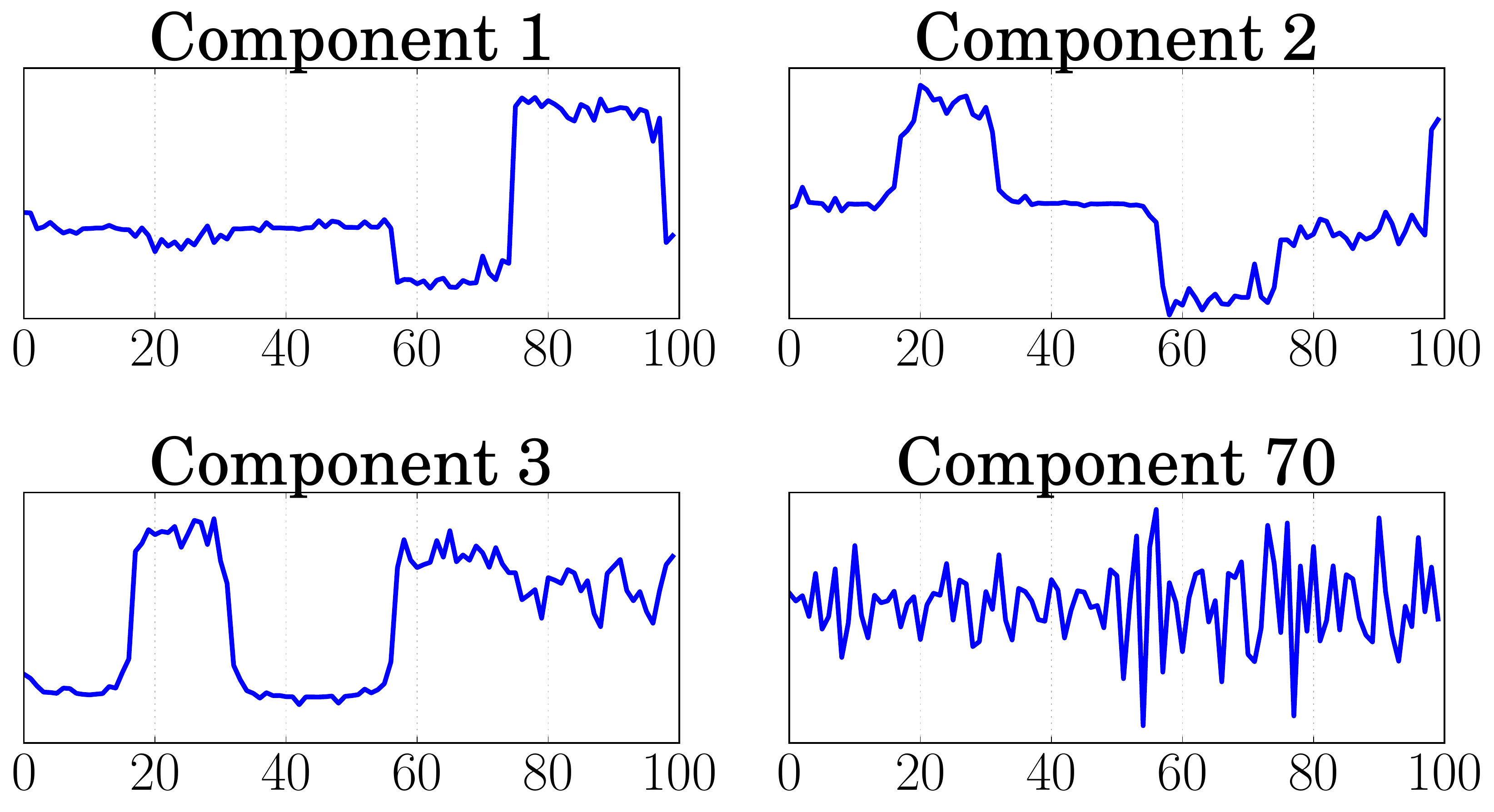}
		\includegraphics[width=0.31\columnwidth]{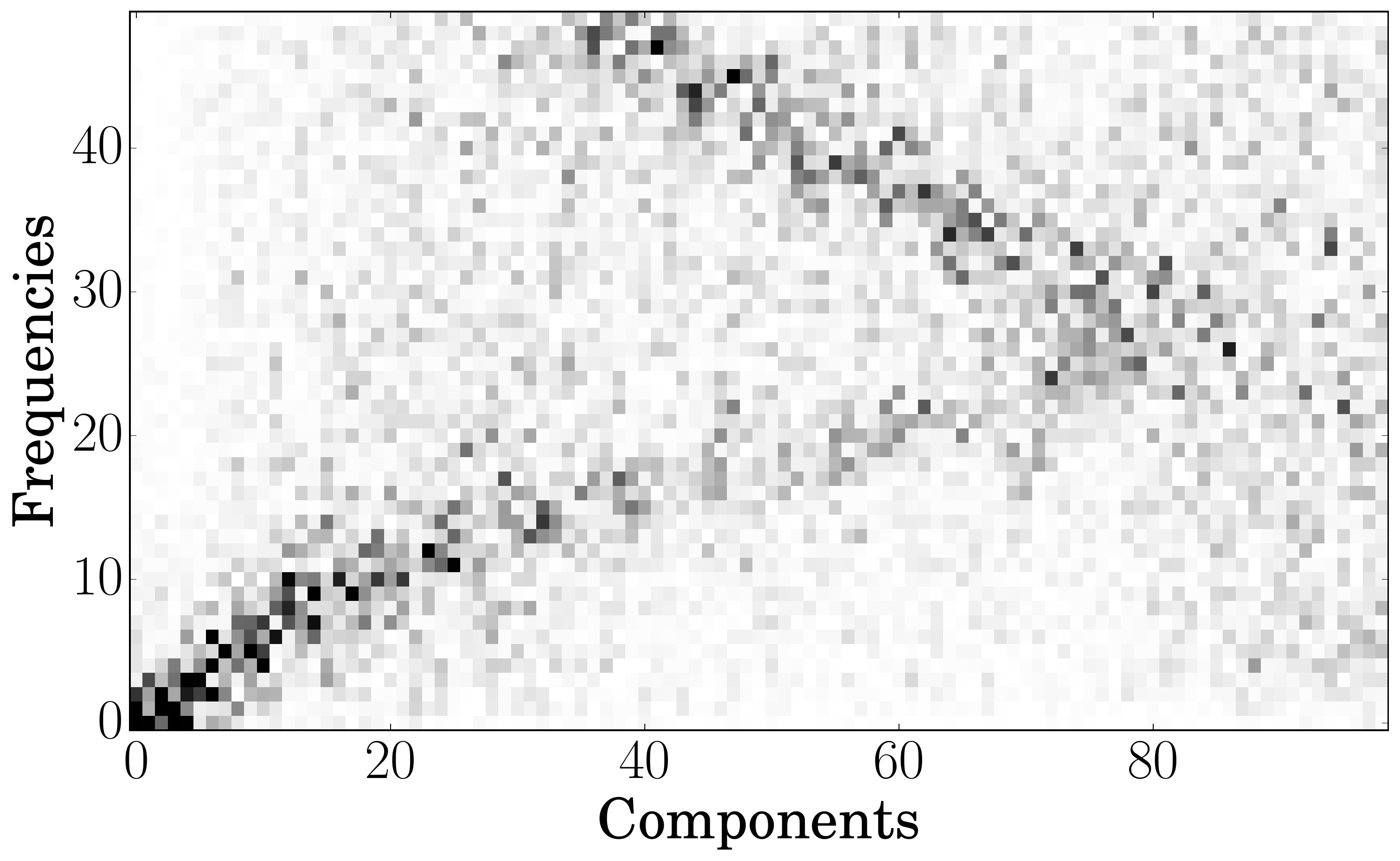}
	}
	
	\subfloat[\label{subfig:gs_er-100-4}Erd\"os-R\'enyi model with $p=0.4$]{
		\includegraphics[width=0.31\columnwidth]{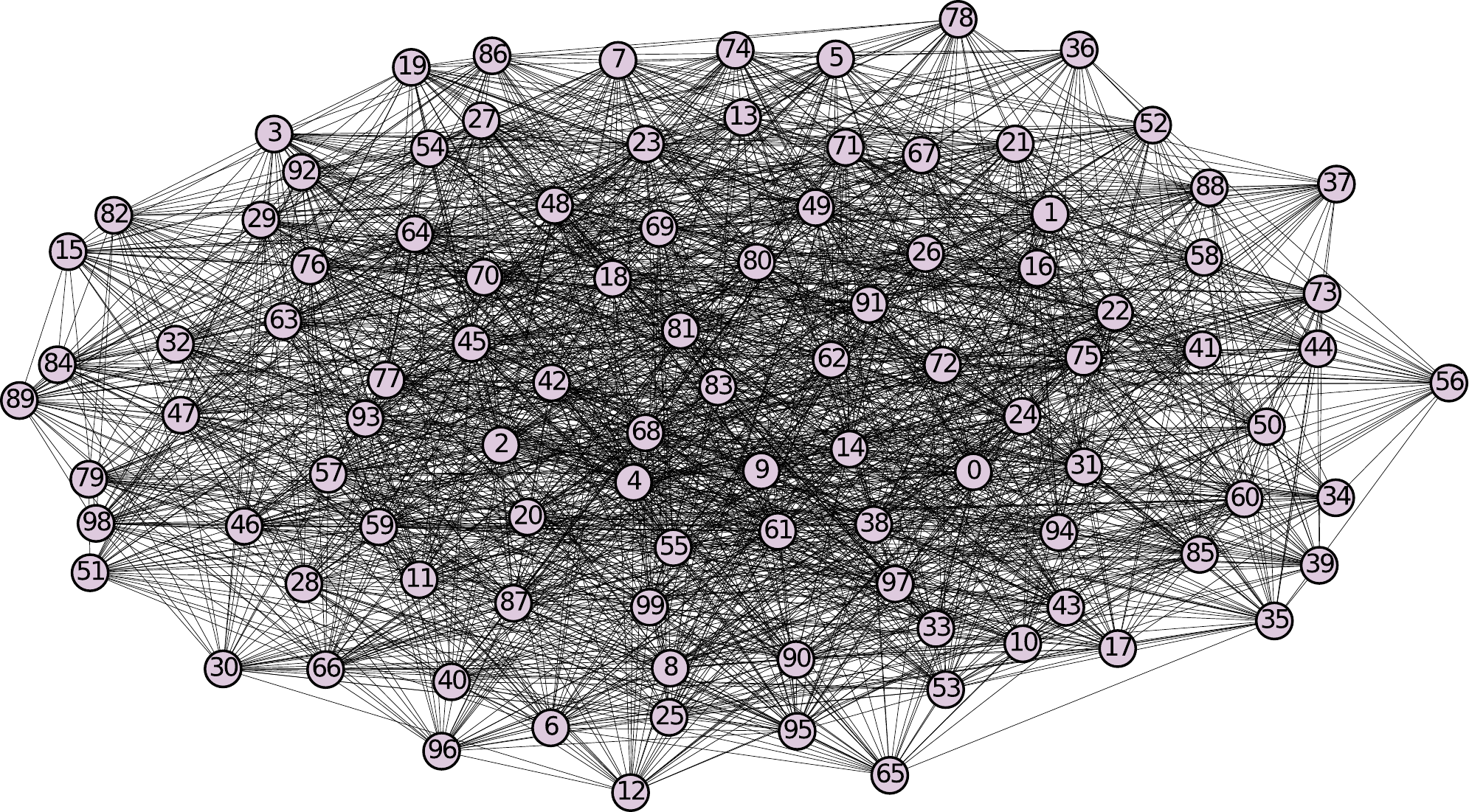}
		\includegraphics[width=0.31\columnwidth]{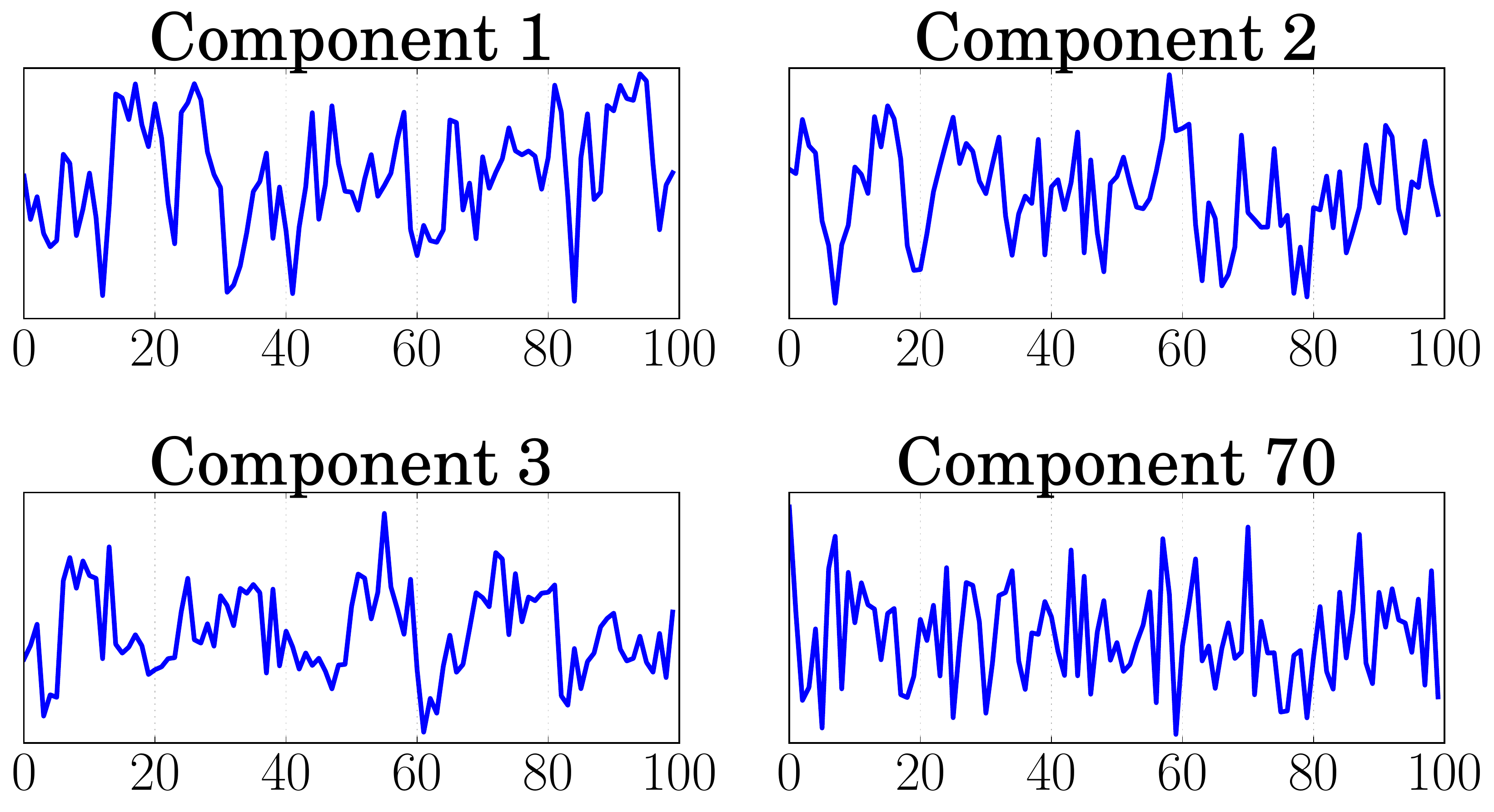}
		\includegraphics[width=0.31\columnwidth]{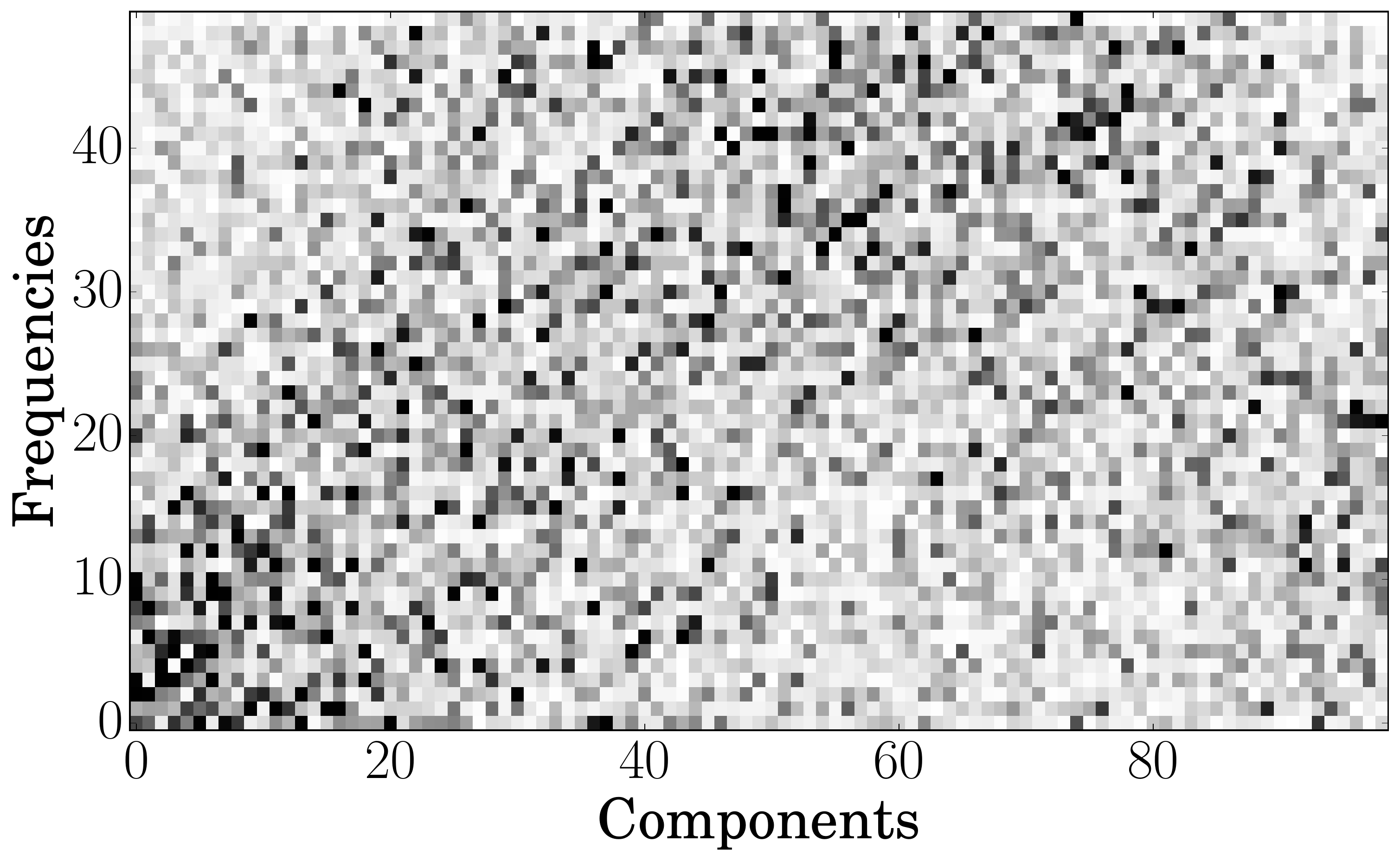}

	}
	
	\caption{\label{fig:GS_illustrations}Illustrations on several instances of 
network models of the transformation of a network into a collection of signals. 
All networks have $N=100$ vertices. (Left) Two-dimensional representation of the 
network. (Middle) First three highest energy components, and the arbitrarily 
chosen component $70$, a low-energy one. (Right) Component-frequency map 
obtained after spectral analysis. The color represents the intensity, coded from 
white to black. The contrast is emphasized to expose the patterns.}

\end{figure}

Figure~\ref{fig:GS_illustrations} shows illustrations of the transformation of a 
graph into a collection of signals, on several instances of network models. 
These illustrations show the connection between graph structure and the 
resulting signals after transformation. Regular structures 
(Figures~\ref{subfig:gs_rl-100-4}~and~\ref{subfig:gs_nrl-100-4-1}) highlight 
harmonic oscillations as components, while an organization in communities 
(Figure~\ref{subfig:gs_sbm-100-3}) presents high-energy frequencies on the first 
components. Combination of both structures in the graph 
(Figure~\ref{subfig:gs_rl_sbm-100-4-3}) is preserved in the frequency pattern. 
Finally, a random graph (Figure~\ref{subfig:gs_er-100-4}) leads to a similar 
absence of structure in the corresponding frequency pattern.


\section{Extension to temporal networks}
\label{sec:extension}

\subsection{Transformation of temporal networks into signals and back}

%
%

The description of temporal networks considered in this work consist in a 
discrete-time sequence of graph snapshots. The collection of all snapshots 
at the different times $t \in \{0,\hdots, T-1 \}$, with $T$ the total 
number of time steps, can be represented by a graph adjacency tensor denoted $\A 
\in \R^{N\times N \times T}$. We study here temporal networks where the edges 
are changing over time, keeping the same given set of vertices (possibly 
isolated).

The extension of the method described in Section~\ref{sec:preliminaries} is 
directly achieved by applying at each time step the transformation on the 
corresponding static representation of the temporal network. We denote $\X \in 
\R^{N\times C \times T}$, the collection of signals obtained from $\A$. For each 
time step $t$, we have
\begin{align}
	\bm{X}^{(t)} = \bm{\T}[\bm{A}^{(t)}]
\end{align}
As the number of vertices in the graph does not evolve, the number of components 
$C$, and then the number of frequencies $F$, is constant over time. In the case 
where the set of vertices evolves, the tensor is built by fixing the number of 
components as the maximal number of components over time, and by zero-padding 
the missing components. As for the indexation of vertices, the algorithm 
relabeling vertices according to the structure of the network is performed at 
each time step, leading to a different labeling of the vertices over 
time.\footnote{The objective of this work here is to track how the structure 
of the temporal network evolves, regardless the labels of the vertices. The 
modification of the labeling indicates how the vertex evolves over time in the 
global structure of the network. This aspect is not discussed in the following, 
the labeling over time is kept only for purposes of reconstruction.}

Conversely, the inverse transformation is performed likewise, by applying the 
static inverse transformation on the collection of signals at time $t$:
\begin{align}
	\bm{A}^{(t, r)} = \bm{\T^{-1}}[\bm{X}^{(t)}]
\end{align}
where $\bm{A}^{(t, r)}$ represents the adjacency matrix of the reconstructed 
temporal network at time $t$. 

\subsection{Illustration on a Toy Temporal Network (TTN)}
\label{subsec:ttn}

\begin{table}

	\caption{\label{tab:probas}Generation of the Toy Temporal Network: 
Probabilities to have an edge at time $t$}
	
	\centering
	\begin{tabular}{|c|c|c|}\hline
		& \textbf{$e \in E^p$} & \textbf{$e \notin E^p$}  \\ \hline
		 \textbf{$e \in E_{t-1}$} & $0.99$ & $0.8$ \\ \hline 
		 \textbf{$e \notin E_{t-1}$}  & $0.2$ & $0.01$ \\ \hline
	\end{tabular}
	
\end{table}

A toy temporal network is introduced for the sake of illustration. The temporal 
network consists of smooth transitions between different network structures. 
Starting from $N$ unconnected vertices, the algorithm adds or removes edges at 
each time instant according to a probability depending both on the presence or 
not of the edge at the previous time, and on a prescribed network structure. Let 
$E_t$ be the set of edges at time $t$, $E^p$ the set of edges of a prescribed 
network structure, and $e = (i,j)$ the edge between the vertices $i$ and $j$. 
Table~\ref{tab:probas} gives the probability to have $e$ in $\E_t$ according to 
the presence of $e$ at the previous time (i.e. $e \in E_{t-1}$) and the presence 
of $e$ in the prescribed network structure ($e \in E^p$). These probabilities 
are set up in order to generate a smooth transition between the initial 
structure and the prescribed network structure, and such that the structure at 
the end of the time interval is close to the one of the prescribed graph.

Four networks are successively chosen as prescribed network structures, among 
those introduced in Section~\ref{sec:preliminaries}, each structure being 
active for a time interval of 20 time steps:
\begin{enumerate}
	\item Random linkage of vertices (see Figure~\ref{subfig:gs_er-100-4})
	\item Network with $3$ communities (see Figure~\ref{subfig:gs_sbm-100-3})
	\item 4-ring lattice (see Figure~\ref{subfig:gs_rl-100-4})
	\item 4-ring lattice with $3$ communities (see 
Figure~\ref{subfig:gs_rl_sbm-100-4-3})
\end{enumerate}

\begin{figure}

	\centering
		\subfloat[\label{subfig:toy_A_time19} $t=19$]{
		\includegraphics[width=0.31\columnwidth]{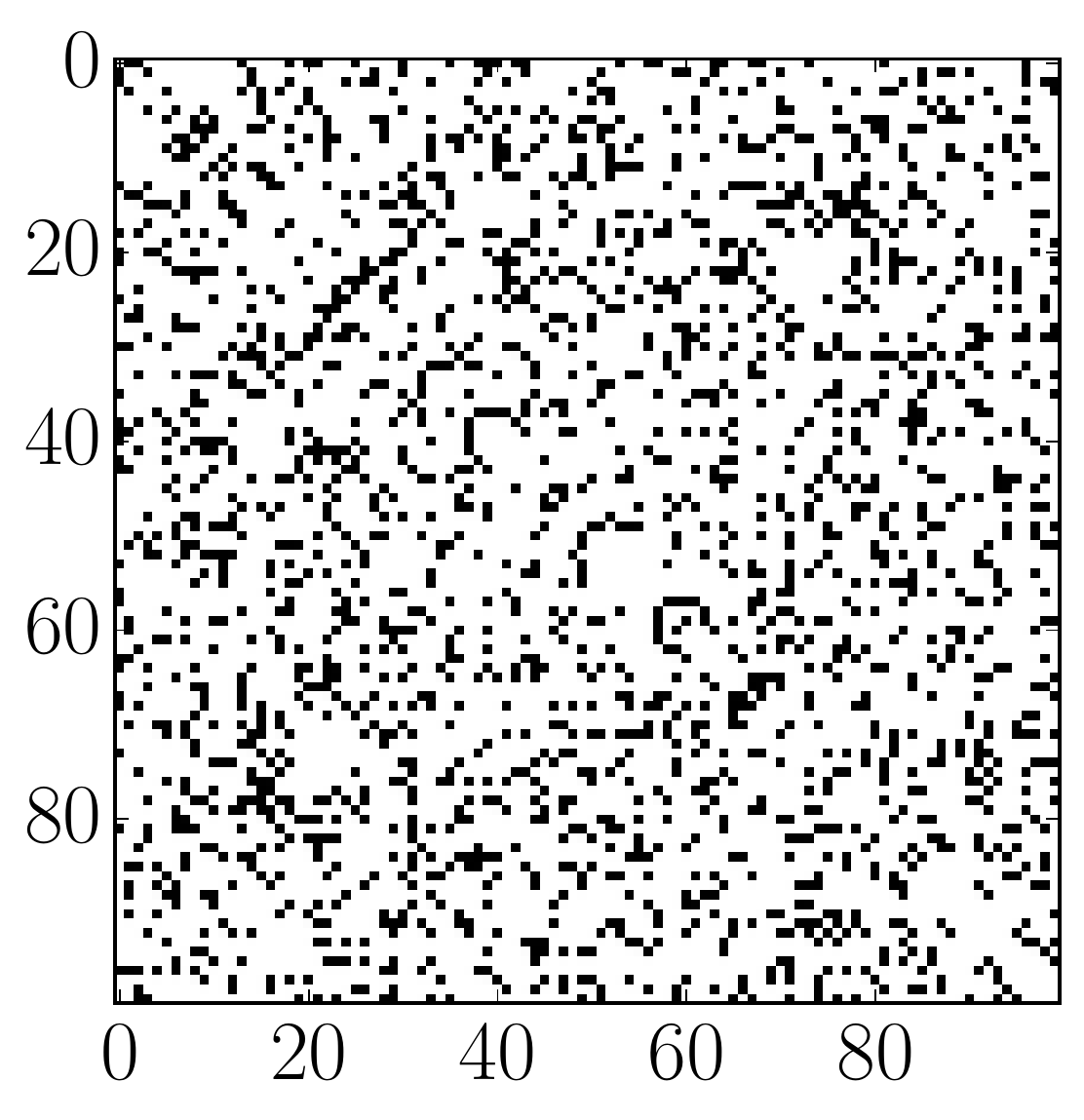}
	}
	\subfloat[\label{subfig:toy_A_time39} $t=39$]{
		\includegraphics[width=0.31\columnwidth]{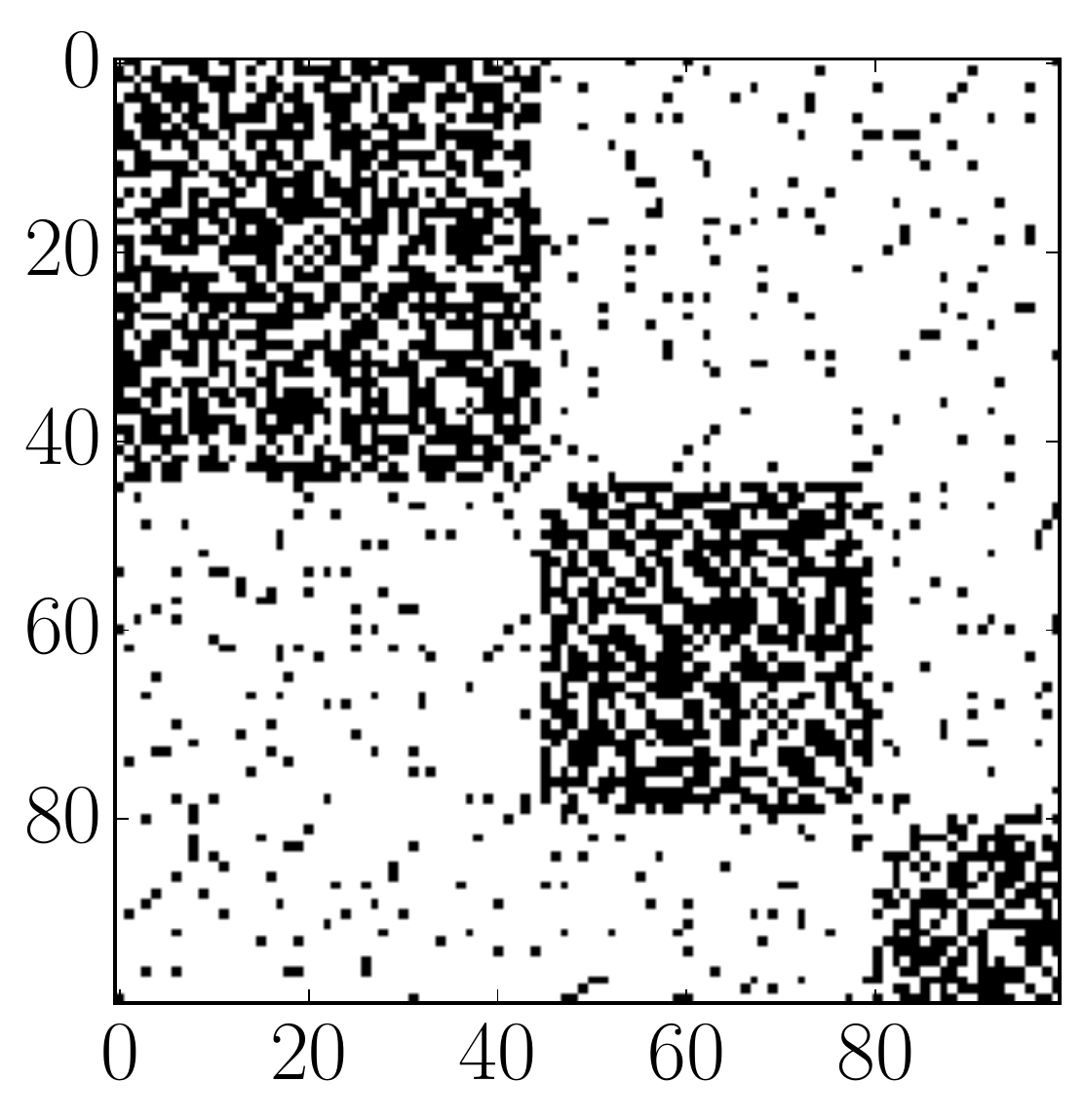}
	}
	\subfloat[\label{subfig:toy_A_time59} $t=59$]{
		\includegraphics[width=0.31\columnwidth]{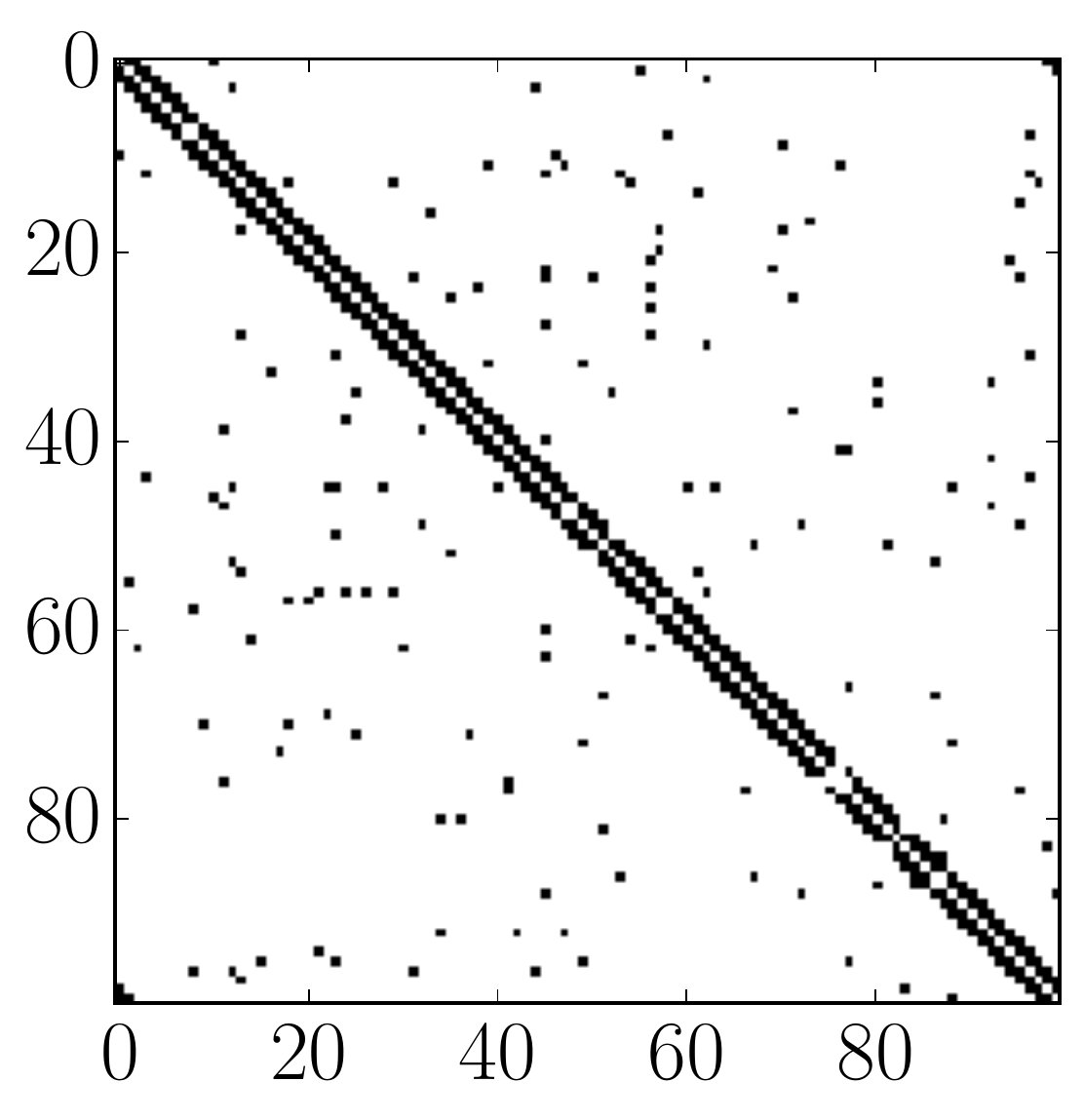}
	}
	
	\subfloat[\label{subfig:toy_A_time79} $t=79$]{
		\includegraphics[width=0.31\columnwidth]{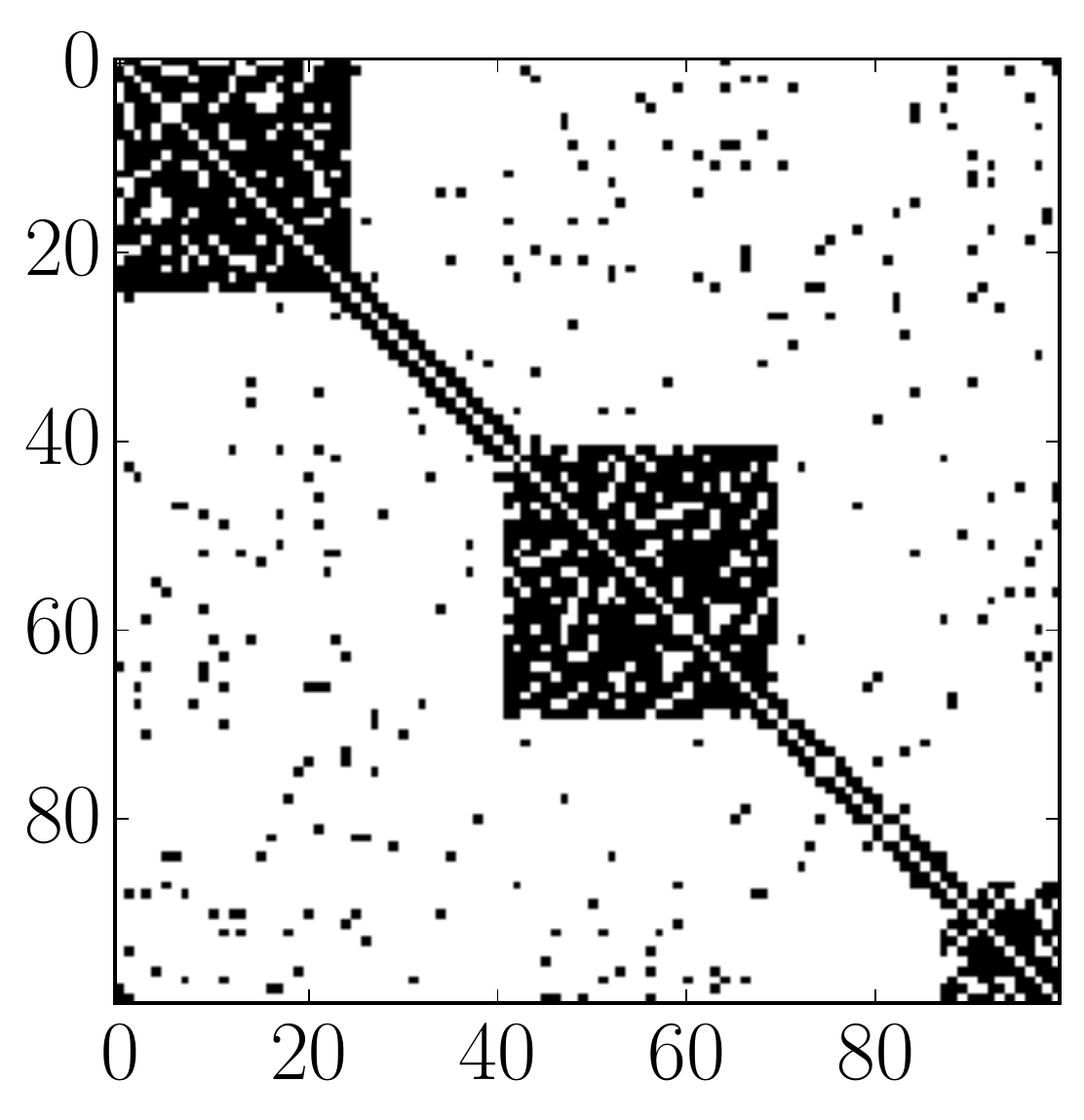}
	}
	\subfloat[\label{subfig:toy_agg_original}Agg. adjacency matrix]{
		\includegraphics[width=0.31\columnwidth]{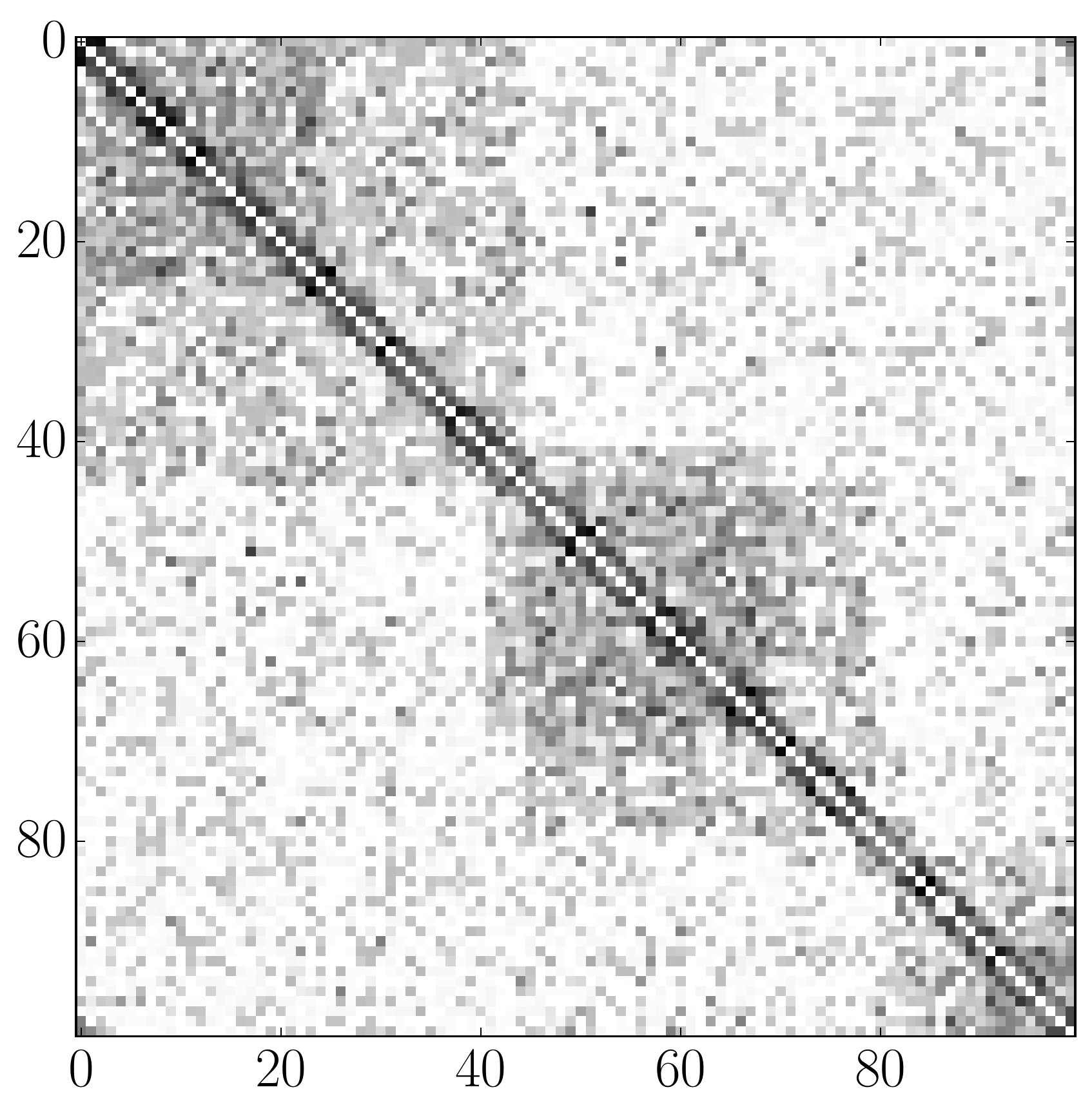}
	}

	\caption{\label{fig:toy_A}Adjacency matrix $\bm{A}_t$ of the Toy 
Temporal Network at different time instant.}

\end{figure}

Figure~\ref{fig:toy_A} shows the adjacency matrix of the temporal network at the 
end of each time interval. The prescribed structure is clearly visible thanks to 
a proper labeling of vertices: the blocks describe the communities, while the 
strong diagonal describes the regular lattice. These structures are nevertheless 
not completely defined, as noise due to the probabilities set in 
Table~\ref{tab:probas} remains present. It allows to assess the robustness of 
the further analyses to noise. Figure~\ref{subfig:toy_agg_original} shows the 
adjacency matrix of the temporal network aggregated over the $80$ time steps: 
the four structures also appear using this representation.

\begin{figure}

	\centering
	\subfloat[\label{subfig:toy_descriptors_nedges} Number of edges]{
		\includegraphics[width=0.95\columnwidth]{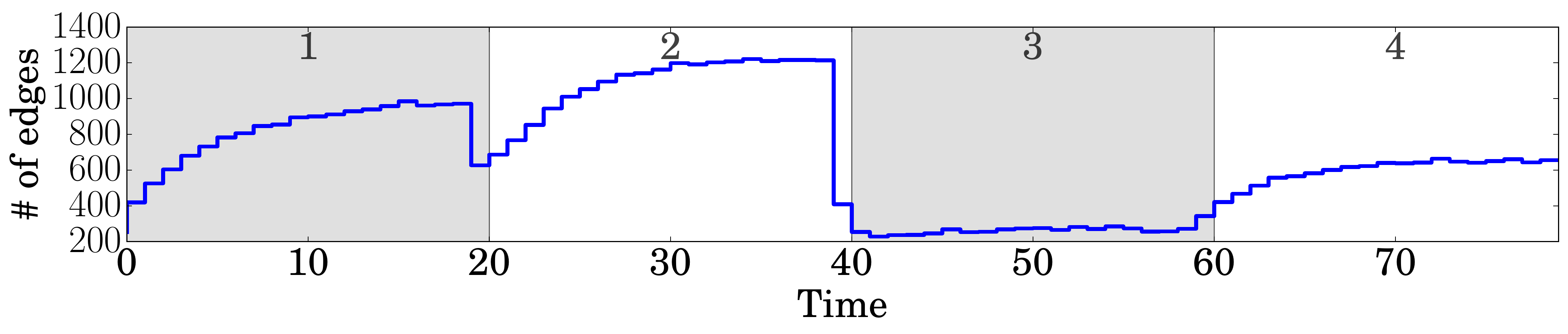}
	}
	
	\subfloat[\label{subfig:toy_descriptors_clust}Average clustering 
coefficient]{
		\includegraphics[width=0.95\columnwidth]{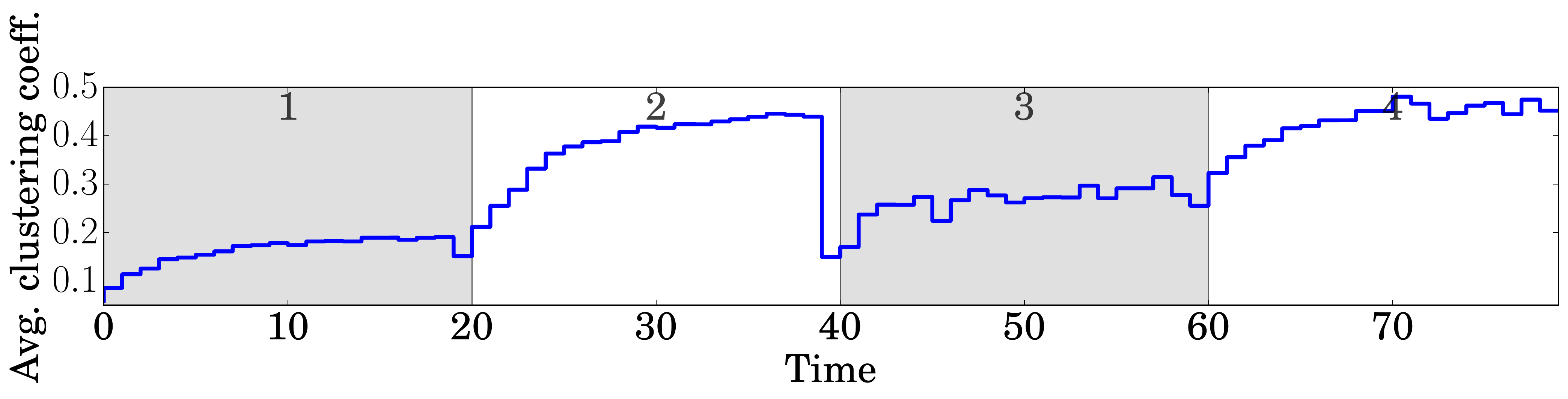}
	}
	
	\subfloat[\label{subfig:toy_descriptors_lensp}Average length of shortest 
paths]{
		\includegraphics[width=0.95\columnwidth]{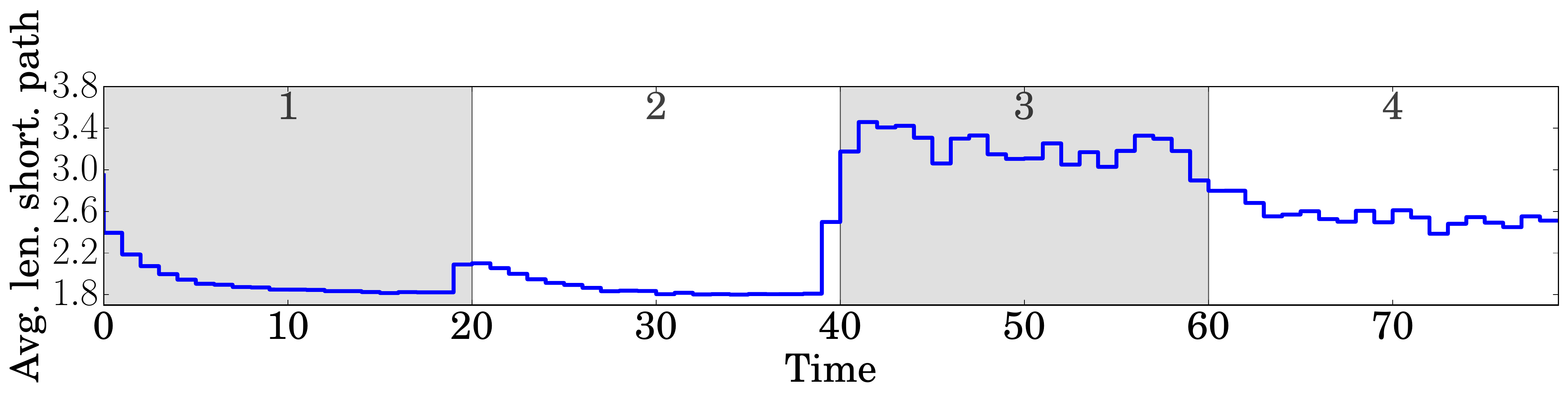}
	}
	
	\caption{\label{fig:toy_descriptors}Descriptors of the Toy Temporal 
Network over time. The alternating shaded regions correspond to the four 
different periods, whose the label is given at the top of each region.}
\end{figure}

Figure~\ref{fig:toy_descriptors} plots three descriptors of the network at each 
time step, which would be a manner to analyze the temporal network using network 
theory. The number of edges (Figure~\ref{subfig:toy_descriptors_nedges}), the 
average clustering coefficient (Figure~\ref{subfig:toy_descriptors_clust}) and 
the average length of shortest paths (Figure~\ref{subfig:toy_descriptors_lensp}) 
allow for an identification of the four time intervals: The number of edges 
increases in the period $2$, to form the random structure, characterized by a 
low average clustering coefficient and low average length of shortest paths. 
Adding communities in period $2$ increases the average clustering coefficient, 
as expected. On the contrary, the regular structure in period $3$ decreases the 
clustering coefficient and increases the average length of shortest paths, but 
much less in proportion than the decrease of the number of edges would affect. 
As expected, adding communities in this regular structure, as it is done in 
period $4$, significantly increases the average clustering coefficient and 
slightly decreases the average length of shortest paths. These network-based 
descriptors give good intuitions on the underlying structure of the temporal 
network, but turn out to be inefficient to exactly characterize the structure. 
Furthermore, mixture of different structures, as it appears in this model, are 
not explicitly revealed. 


\section{Spectral Analysis of Temporal Networks}
\label{sec:spectral}

\subsection{Temporal Spectra}

As previously, the extension of the spectral analysis of signals representing 
networks, defined in Section~\ref{subsec:spectral_analysis} is simply achieved 
by considering independently each time step. We denote by $\Sc \in \R^{C \times 
F\times T}$ the spectral tensor, where $\bm{S}^{(t)}$ corresponds to the spectra 
obtained at time $t$. The corresponding matrices $\M$ and $\E$ are respectively 
the temporal magnitudes and energies.

As described in Section~\ref{sec:preliminaries}, the spectra are closely related 
to the network structures. In particular, the importance of high-energy 
components as well as low frequencies have been highlighted, for instance for 
the structure in communities. Looking at the marginals of the temporal energies 
or magnitudes over frequencies and components is hence expected to give hints 
about the evolution of the structure of the temporal network over time. In the 
following, we will focus on the marginal of the energies over the components, 
denoted $\bm{E}_c(f)$, and over the frequencies, denoted $\bm{E}_f(c)$.

We also propose to use the spectral analysis to compare two network structures 
by computing the correlation between their spectra at each time step. The idea 
is to find among a set of parametric graph models the one that best fit the 
network at each time step. If we denote by $\bm{S}_m$ the spectra obtained after 
transformation of an instance $\G_m$ of a prescribed graph model, we can compute 
a correlation coefficient $\rho^{(t)}$ between $\bm{S}_m$ and $\bm{S}^{(t)}$. 
Generating several instances of the graph model gives us an average value of the 
correlation coefficient over several repetitions.

\subsection{Illustration on the Toy Temporal Network (TTN)}

\begin{figure}

	\centering
	
	\subfloat[\label{subfig:toy_marginals_freq}Averaged over frequencies 
$\bm{E}_f(c)$ for $c \in 
\{0, 1, 2, 70\}$]{	
		\includegraphics[width=0.95\columnwidth]{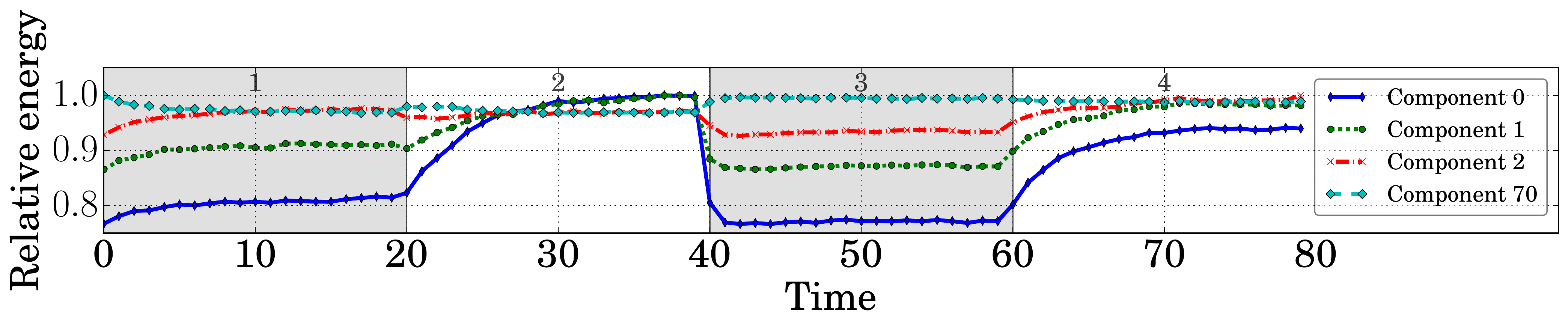}
	}
	
	\subfloat[\label{subfig:toy_marginals_comp}Averaged over components 
$\bm{E}_c(f)$ for $f \in 
\{1, 2, 3, 35\}$]{	
		\includegraphics[width=0.95\columnwidth]{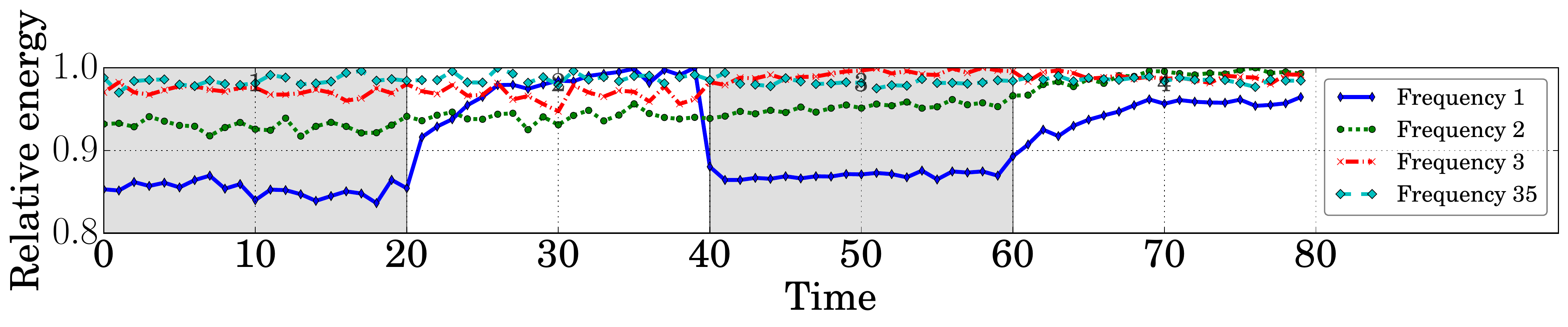}
	}
	
	\caption{\label{fig:toy_marginals}Marginals of the energies of 
temporal spectra.}

\end{figure}

Figure~\ref{fig:toy_marginals} shows $\bm{E}_c(f)$ and $\bm{E}_f(c)$ for 
respectively $f \in \{1, 2, 3, 35\}$ and $c \in \{0, 1, 2, 70\}$. These two 
figures reveal the predominance of the first components and low frequencies,  to 
track an organization in communities of the network: the low frequencies for the 
first components have a greater energies than for other types of structure, as 
already remarked in Section~\ref{subsec:illustrations}. The tracking of the 
other types of structures is nevertheless not visible using this representation.

\begin{figure}

	\centering
	
	\subfloat[\label{subfig:toy_corr_com}Comparison with a network with $k$ 
communities, for different values of $k$. The temporal network is highly 
correlated with a graph organized in communities during the period $2$ and $4$, 
as expected.]{	
		\includegraphics[width=0.95\columnwidth]{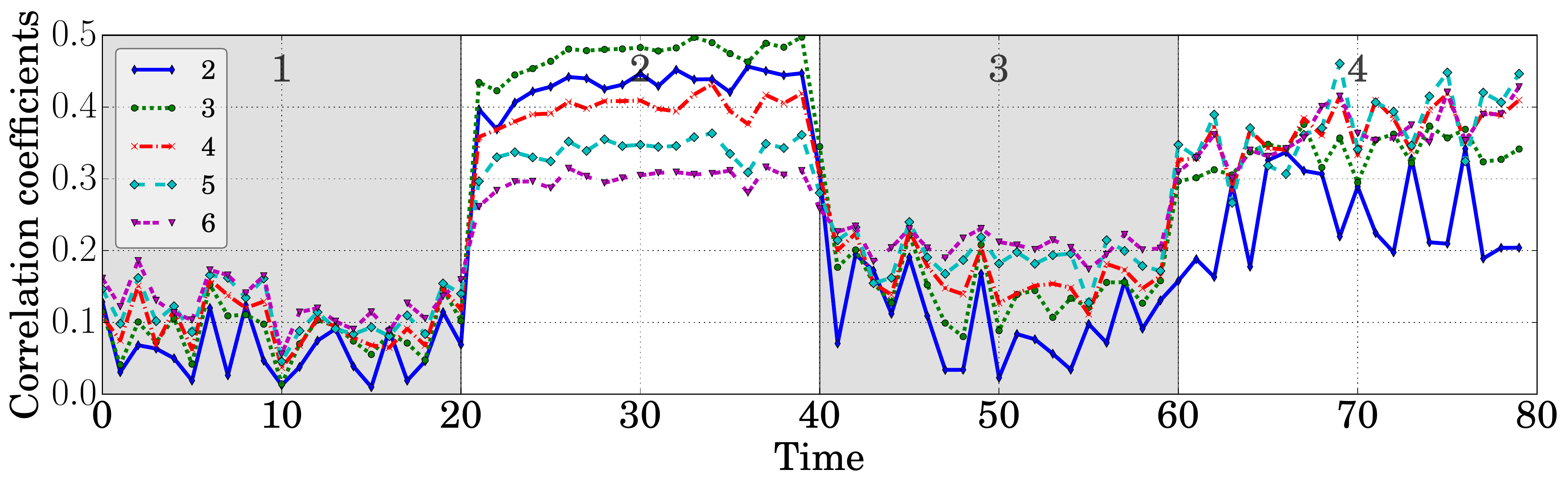}
	}
	
	\subfloat[\label{subfig:toy_corr_ws}$k$-regular lattice, for different value 
of $k$. The temporal network has the highest correlation with a $4$-regular 
lattice during the period $3$.]{	
		\includegraphics[width=0.95\columnwidth]{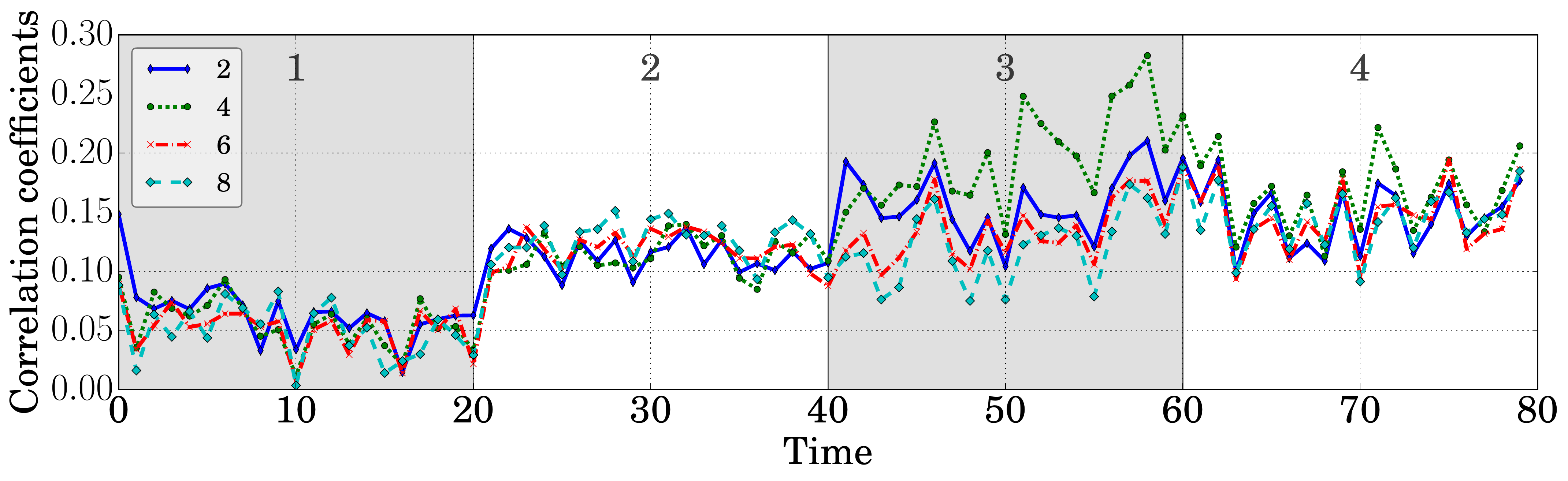}
	}
	
	\caption{\label{fig:toy_corr}Correlation between the temporal spectra at 
each time step and the spectra of networks with two specific network structures. 
The correlation is averaged over $20$ repetitions.}

\end{figure}

The correlation between the temporal spectra of the toy temporal network and two 
network structures is studied. First, a structure in communities is observed, 
using a network model generating a random graph with a fixed number of 
communities. The comparison is done for a number of communities from $2$ to $6$, 
with $20$ repetitions for each number of communities. 
Figure~\ref{subfig:toy_corr_com} shows the average correlation: the correlation 
is maximal during the periods $2$ and $4$, where the network is effectively 
structured in communities. During the period $2$, the correlation is maximal 
when the network is structured in $3$ communities, as expected. During the 
period $4$, the number of communities is not clearly revealed, as the 
communities are not the only component of the network topology.

In Figure~\ref{subfig:toy_corr_ws}, the temporal network is compared using the 
same method with a $k$-regular lattice, for $k$ equals to $2$, $4$, $6$ and $8$. 
If the correlations are lower than previously, we can nevertheless notice that 
in period $3$, the temporal network is correlated with a $4$-regular lattice, 
which is the structure set in the prescribed network during this period.

The study of temporal spectra give hence insights about the structure of the 
underlying temporal network, but this approach is limited by the lack of 
knowledge, in the general case, of the parametric graph models composing the 
temporal network. To complement our analysis, we propose in the following to 
automatically find structures and their corresponding time activation periods 
that best characterize the temporal network.

\section{Extraction of Frequency Patterns using Nonnegative Matrix 
Factorization}
\label{sec:decomposition}

\subsection{Nonnegative Matrix Factorization (NMF)}

Nonnegative matrix factorization (NMF) \cite{Lee1999} is a linear regression 
technique, used to decompose a nonnegative matrix $\bm{V}$  of dimension $C 
\times T$, i.e. a matrix whose terms are greater or equal to zero, into the 
product of two nonnegative matrices $\bm{W}$ and $\bm{H}$. NMF leads to a 
reduction of the dimensionality of data, by extracting in the columns of 
$\bm{W}$ patterns characterizing the data, and in the rows of $\bm{H}$ the 
activation coefficients of each pattern along the time. The number of extracted 
patterns is denoted $K$. A common approach to achieve such a decomposition 
consists of solving an optimization problem:
\begin{align}
	\label{eq:nmf}
	(\bm{W}^*, \bm{H}^*) = \arg\min_{\bm{W}\bm{H}} D(\bm{V} | \bm{WH})
\end{align}
where $D$ is a dissimilarity measure. F\'evotte et al. \cite{Fevotte2011b} 
proposed an algorithm to find a solution of the NMF where $D$ is the 
$\beta$-divergence, a parametric function with a simple parameter $\beta$ 
which encompasses the Euclidean distance ($\beta=2$), the generalized 
Kullback-Leibler divergence ($\beta=1$) and the Itakura-Saito divergence 
($\beta=0$) as special cases. 

Regularization of the activation coefficients can be added in order to smooth 
them, with the assumption that there is no abrupt variations in the structure 
from one time step to the next one. The optimization problem is then defined as 
the minimization of the fitting term, defined in Eq.~\ref{eq:nmf}, plus a term 
of temporal regularization:
\begin{align}
	P(\bm{H}) =  \sum_{k=1}^K\sum_{t=2}^T D(h_{k(t-1)} | h_{kt})
\end{align}
leading to:
\begin{align}
	\label{eq:nmf_regularized}
	(\bm{W}^*, \bm{H}^*) = \arg\min_{\bm{W}\bm{H}} D(\bm{V} | \bm{WH}) + 
\gamma P(\bm{H})
\end{align}
where $\gamma$ controls the regularization and is empirically fixed such that 
the activation coefficients highlight smoothness. In \cite{Fevotte2011a} and 
\cite{Essid2013}, smooth NMF has been introduced for $\beta = 1$ and $\beta=0$.

\subsection{NMF on Spectra of Graphs}

Several approaches have been proposed to adapt NMF to networks, either static 
\cite{Zhang2007} or temporal \cite{Gauvin2014}. In the latter approach, the 
adjacency matrix is represented as a tensor and is decomposed using nonnegative 
tensor factorization (NTF) \cite{Cichocki2014}. The drawback of this approach is 
that the adjacency matrix at each time step is represented as the product of 
vectors, which is well-suited to highlight structure in communities but not 
adapted when the structure becomes more complex.

Following \cite{Hamon2014a}, we propose to use NMF to find patterns in spectra 
of the collections of signals, obtained from the transformation of the temporal 
network. By analogy with music analysis, where an audio sample is decomposed 
into several audio samples, separating for instance voice from the instrumental 
part \cite{Fevotte2009}, we would like to decompose the temporal network into 
temporal sub-networks, decomposing at each time step the global structure into 
several substructures. Furthermore, audio spectra share similarities with graph 
spectra. We then propose to use the Itakura-Saito divergence as measure of 
dissimilarity, given by $d_{IS}(x|y) = \frac{x}{y} - \log\frac{x}{y} - 1$. It 
allows taking advantage of the framework developed to reconstruct the signals 
from spectra as well as using efficient optimization algorithms with possible 
temporal regularization.

As the input in our case is the temporal spectra $\Sc$, represented as a tensor 
of dimension $C \times F \times T$, a small adaptation has to be performed 
before applying NMF. At each time instant $t$, the spectra $\bm{S}_t$ is 
represented as a vector $\bm{v}_t$ by successively adding end-to-end the columns 
of the matrix $\bm{S}_t$. For all $t \in \{0, \hdots, T-1 \}$, these vectors 
compose the columns of the matrix $\bm{V}$, of dimension $(FC) \times T$. The 
number of components $K$ is set according to our expectations about the data, 
and the parameter $\gamma$ is strictly positive to ensure smoothness in the 
activation coefficients. 

\subsection{Identification of components}

NMF returns two matrices $\bm{W}$ and $\bm{H}$: each column of $\bm{W}$ 
represents the $k$th (normalized) frequency pattern, while the $k$th column of 
$\bm{H}^T$, gives the activation coefficients of the frequency pattern $k$ at 
each time step. From $\bm{w}_k$, a component-frequency map can be built by 
reshaping the vector into a matrix. To highlight how these structures are 
arranged in the temporal network, each component is transformed into temporal 
network: As described in \cite{Fevotte2009}, using NMF with the Itakura-Saito 
divergence provides means of reconstruction of the collection of signals 
corresponding to each component. For each component $k$, a temporal spectrum 
$\Sc^{(k)} \in \R^{C \times F\times T}$ is obtained using a Wiener filtering 
such that its elements $s_{cf}^{(k, t)}$ read as:
\begin{align}
	s^{(k, t)}_{cf} = \frac{w_{(cf)k} h_{kt}}{\sum_{l=1}^K w_{(cf)l} h_{lt}} 
s_{cf}
\end{align}
leading to a conservative decomposition of the tensor $\Sc$:
\begin{align}
	\Sc = \sum_{k=1}^K  \Sc^{(k)}
\end{align}
The temporal spectrum of the component $k$ is then a fraction of the original 
temporal spectrum. From $\Sc^{(k)}$, an inverse Fourier transformation is 
performed, leading to a collection of signals for each component $k$ denoted by 
$\X^{(k)} \in \R^{N\times N\times T}$. Finally, the adjacency tensor $\A^{(k)}$ 
describing the temporal network corresponding to the component $k$ is obtained 
by using the inverse transformation $\bm{\T^{-1}}$ described in 
Section~\ref{sec:extension}.

\subsection{Application to the Toy Temporal Network (TTN)}
\label{subsec:nmf_ttn}

\begin{figure}

	\centering
	\subfloat[\label{subfig:toy_comp_w}Frequency patterns, obtained after 
reshaping the columns of $\bm{W}$ into matrices.]{
		\includegraphics[width=0.95\columnwidth]{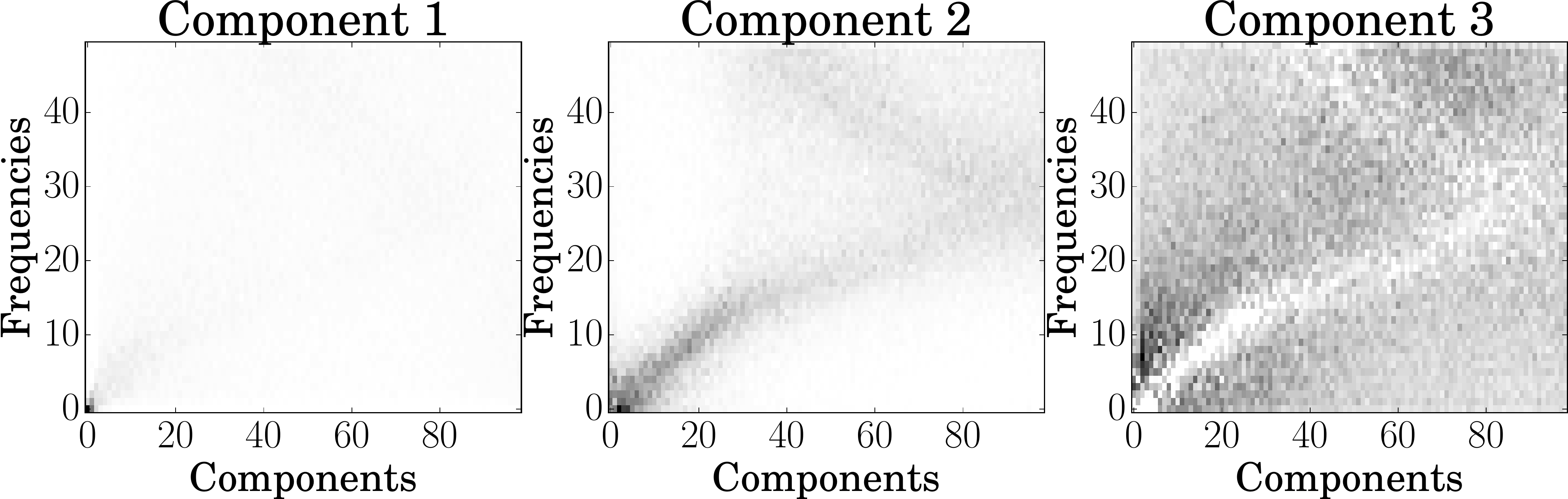}
	}
	
	\subfloat[\label{subfig:toy_comp_h}Activation coefficients, corresponding to 
the rows of the matrix $\bm{H}$, normalized by the maximal value of $\bm{H}$]{
		\includegraphics[width=0.95\columnwidth]{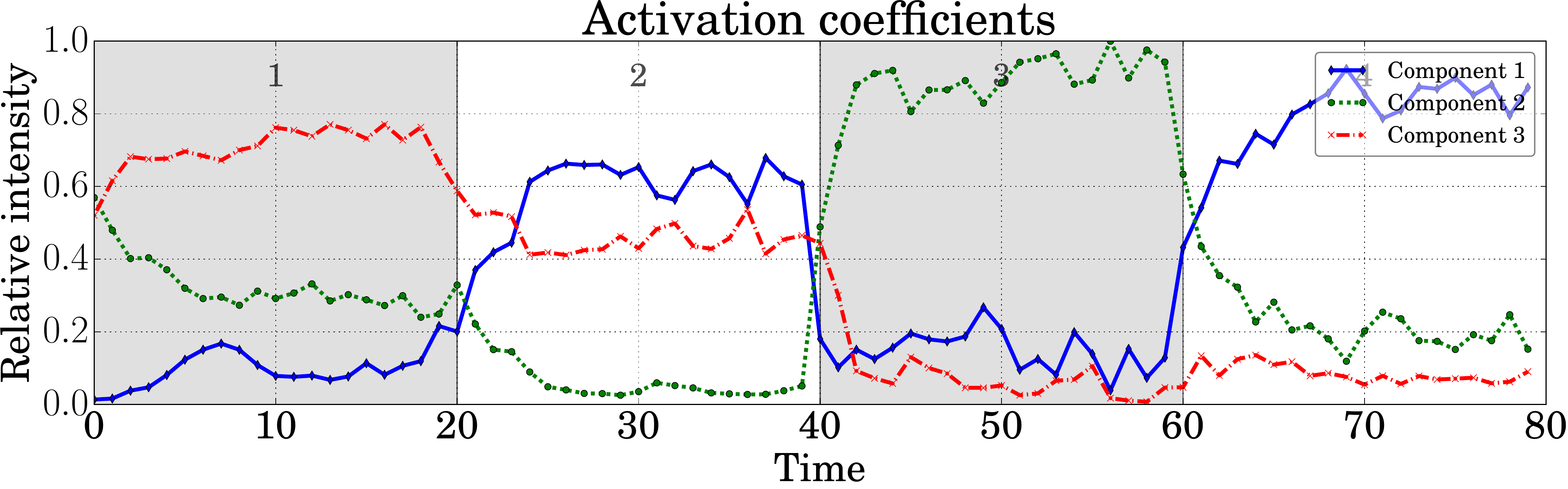}
	}
	\caption{\label{fig:toy_comp_wh}Results of the nonnegative matrix 
factorization for the Toy Temporal Network, using $K=3$ and $\gamma=5$.}

\end{figure}

NMF is applied to the Toy Temporal Network defined in Section~\ref{subsec:ttn}. 
The matrix $\bm{V} \in \R^{4950 \times 80}$ is decomposed using $K=3$ and 
$\gamma=5$, which is the expected number of structures. 
Figure~\ref{subfig:toy_comp_h} shows the activation coefficients of each 
component. We notice that the activation coefficients are consistent with the 
division of time introduced in the generation of the temporal network: All 
components have  distinct levels of activation, corresponding to the four 
different structures used. The component $1$ is active in periods $1$, $2$ and 
$4$ with an almost constant level, the component $2$ is mainly active in period 
$3$, as well as in periods $1$ and $4$, and finally, the component $3$ is active 
in period $1$ and in period $2$. 

Looking at the corresponding frequency patterns in 
Figure~\ref{subfig:toy_comp_w}, and more precisely their connections with the 
network structures observed in Section~\ref{subsec:illustrations}, confirms the 
good adequacy with the expected results: The component $1$ looks like a 
structure in communities, the component $2$ resembles a $k$-regular structure 
and the component $3$ exhibits random structure. The structure in period $1$ is 
then a mixture between a random structure and a structure in communities (as it 
happens, one single community), in period $2$ only the structure in communities 
is present, in period $3$ a regular structure described by component $2$ and 
finally, the fourth period is a mixture between structure in communities and 
regular structure. Random structure is present in period $1$ and in period $2$ 
inside communities.

\begin{figure}

	\centering	
	\subfloat[\label{subfig:toy_agg_comp1}Component 1]{
		\includegraphics[width=0.31\columnwidth]{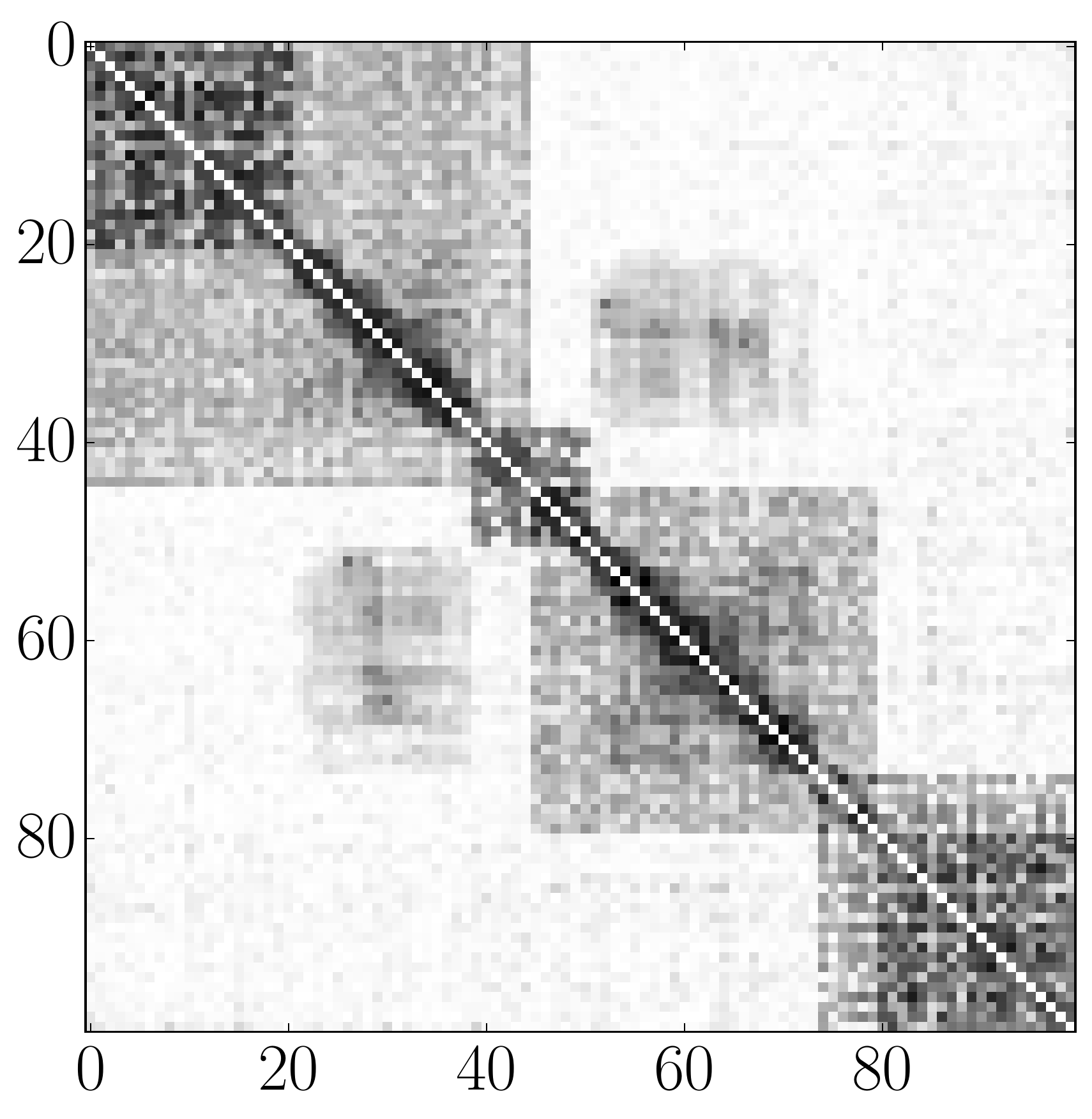}
	}
	\subfloat[\label{subfig:toy_agg_comp2}Component 2]{
		\includegraphics[width=0.31\columnwidth]{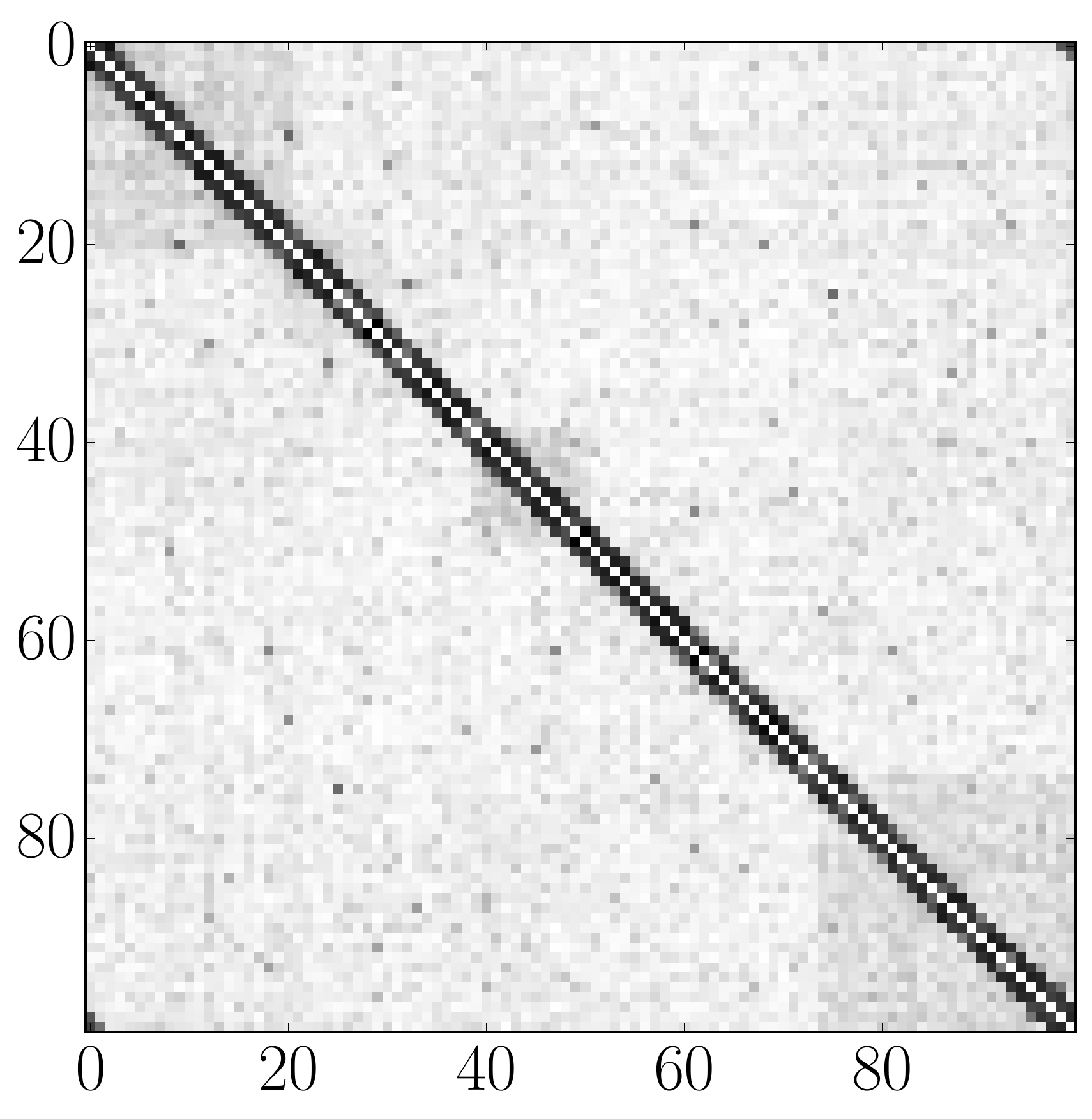}
	}
	\subfloat[\label{subfig:toy_agg_comp3}Component 3]{
		\includegraphics[width=0.31\columnwidth]{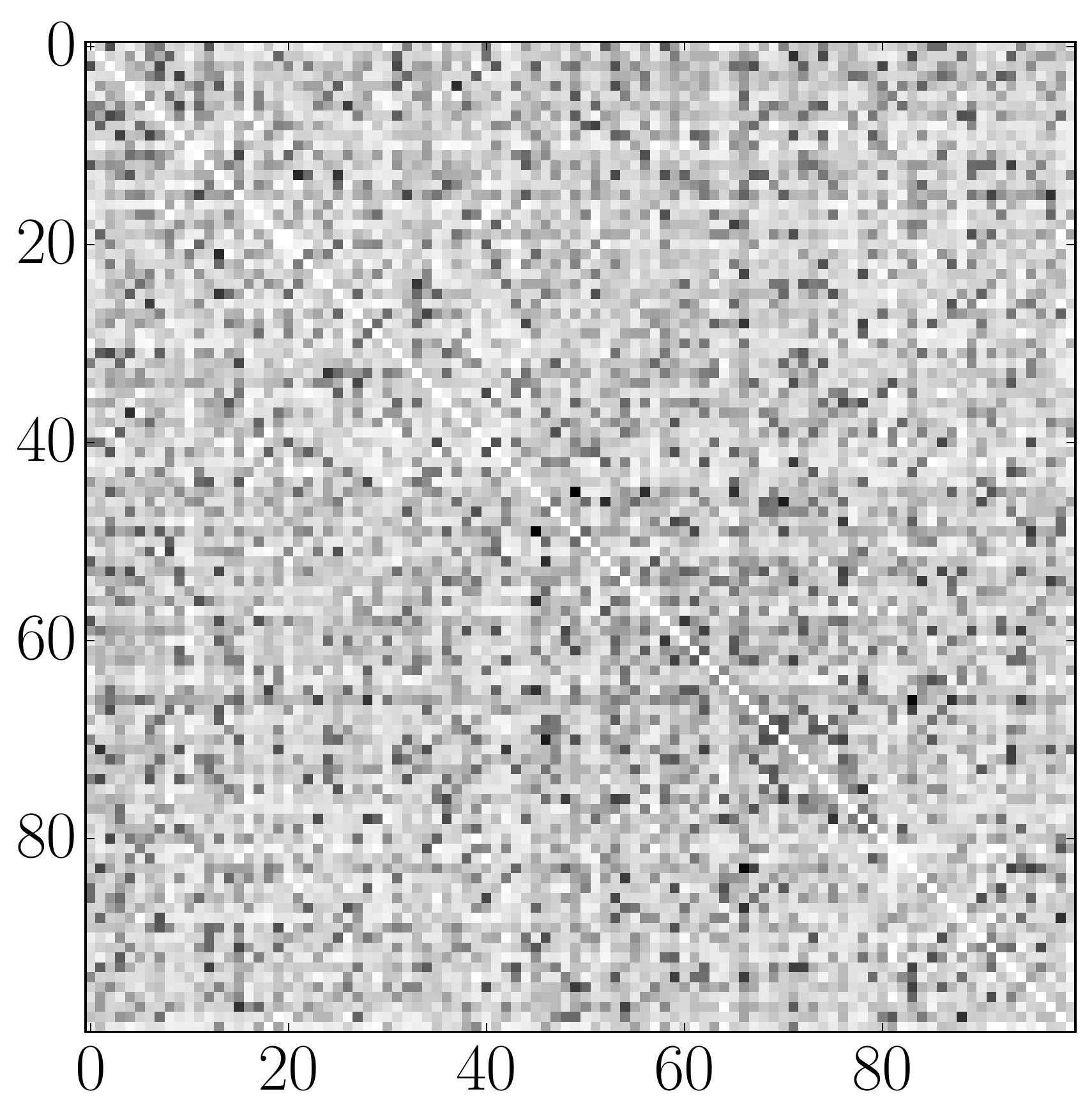}
	}

	\caption{\label{fig:toy_agg}Adjacency tensor aggregated over time: 
for each component $k$, $\bm{A}^{(k)} = \sum_{t=1}^T h_{kt} \bm{A}^{(k, t)}$.}
\end{figure}

Figure~\ref{fig:toy_agg} shows, for each component $k$, the aggregated adjacency 
matrix over time, obtained after the back transformation of the spectra into a
temporal network. The sum is weighted by the activation coefficients given by 
the matrix $\bm{H}$, in order to highlight visually the most significant 
patterns: 
\begin{align}
	\bm{A}^{(k)} = \sum_{t=1}^T h_{kt} \bm{A}^{(k, t)}
\end{align}
This representation partly confirms the connections between spectra and 
structures as described above. We can notice that component $1$ displays the 
three communities present in period $2$, as well as communities corresponding to 
a regular structure in period $4$. The actual communities in period $4$ are 
caught in the component $2$, which does not correspond to a regular structure, 
even if the diagonal, characterizing the $k$-regular lattice, is clearly 
dominant. Finally, the component $3$ looks like a random matrix, as no structure 
is visible, at least through this representation.

The decomposition of temporal networks using NMF enables to retrieve from the 
spectra the different structures composing the temporal network, and to detect 
when these structures are present, either alone or jointly. We now propose to 
apply this framework to study a real-world temporal network describing contacts 
between individuals in a primary school.

\section{Temporal Network of Social Interactions Between Children in a 
Primary School}
\label{sec:primary}

\subsection{Description of the temporal network}

The decomposition is applied to a real-world temporal network, describing social 
interactions between children in a primary school during two days in October 
2009. During a 20-second interval, an edge exists between two individuals if a 
contact is recorded, measured by wearable RFID (Radio Frequency IDentification) 
sensors \cite{Stehle2011}. For our study, the data set is described by a 
temporal network, representing for each time step the aggregated contacts 
between individuals for 10 minutes. We restrained the analysis to the first day: 
226 children and 10 teachers participated in the experiment, separated in five 
grades (from 1st grade to 5th grade), themselves separated in two classes.

\begin{figure}

	\centering
	\subfloat[\label{subfig:prim600_descriptors_nedges} Number of edges]{
		\includegraphics[width=0.95\columnwidth]{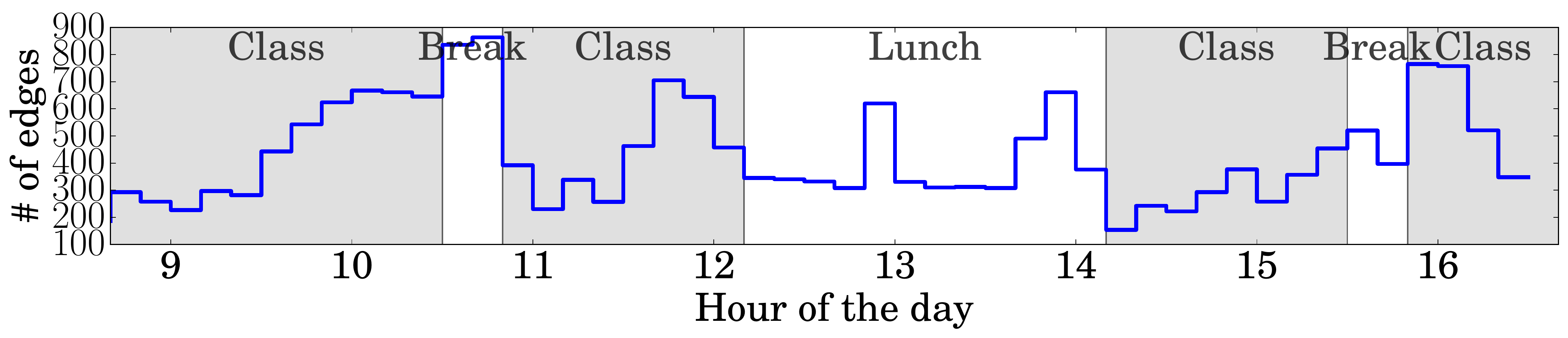}
	}
	
	\subfloat[\label{subfig:prim600_descriptors_isol}Number of isolated 
vertices]{
		\includegraphics[width=0.95\columnwidth]{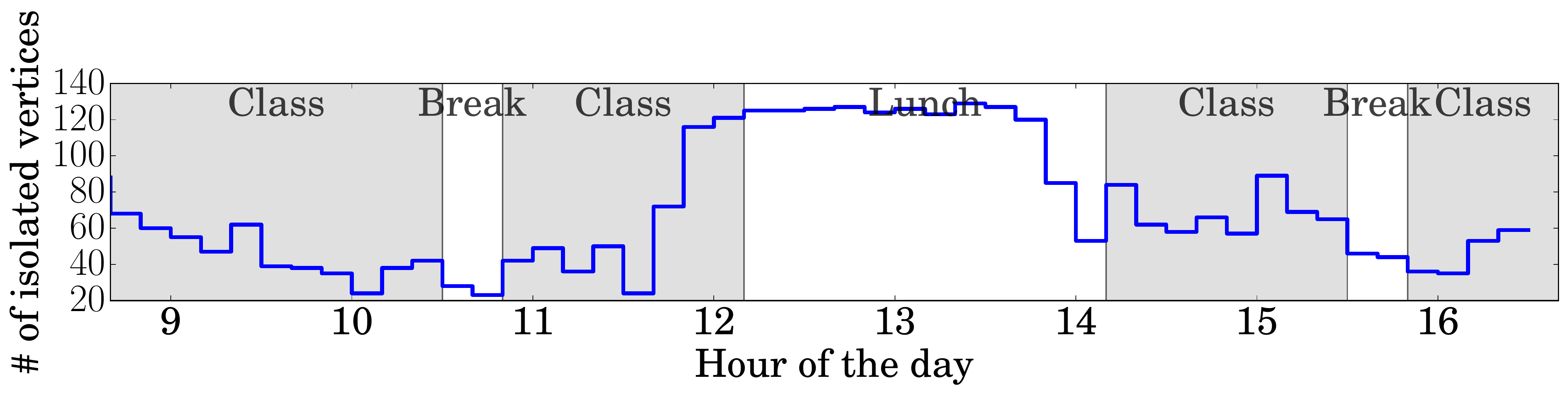}
	}
	
	\subfloat[\label{subfig:prim600_descriptors_clust}Average clustering 
coefficient]{
		\includegraphics[width=0.95\columnwidth]{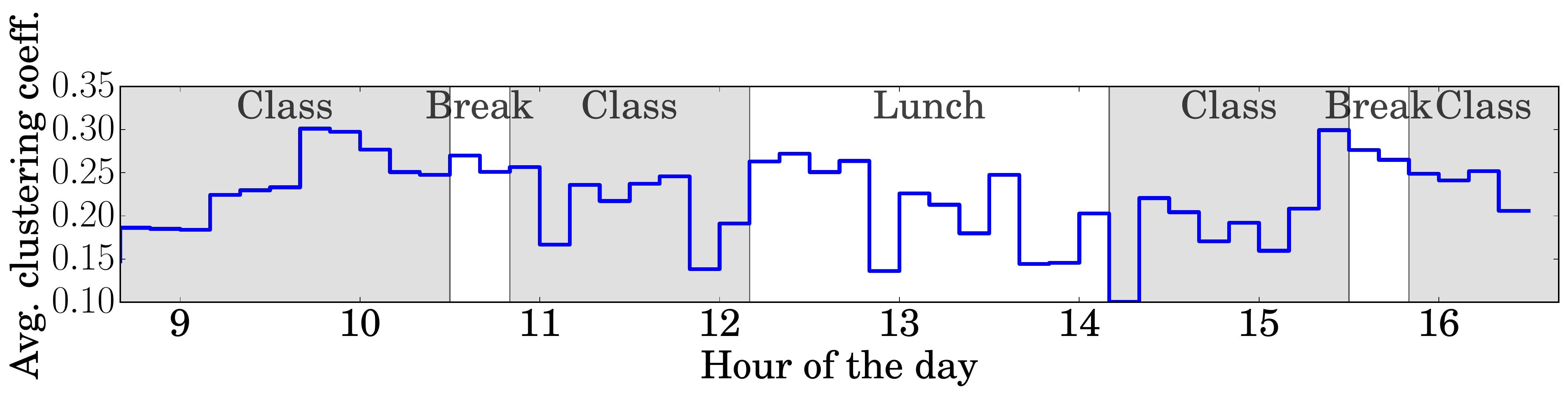}
	}
	
	\subfloat[\label{subfig:prim600_descriptors_lensp}Average length of 
shortest 
paths]{
		\includegraphics[width=0.95\columnwidth]{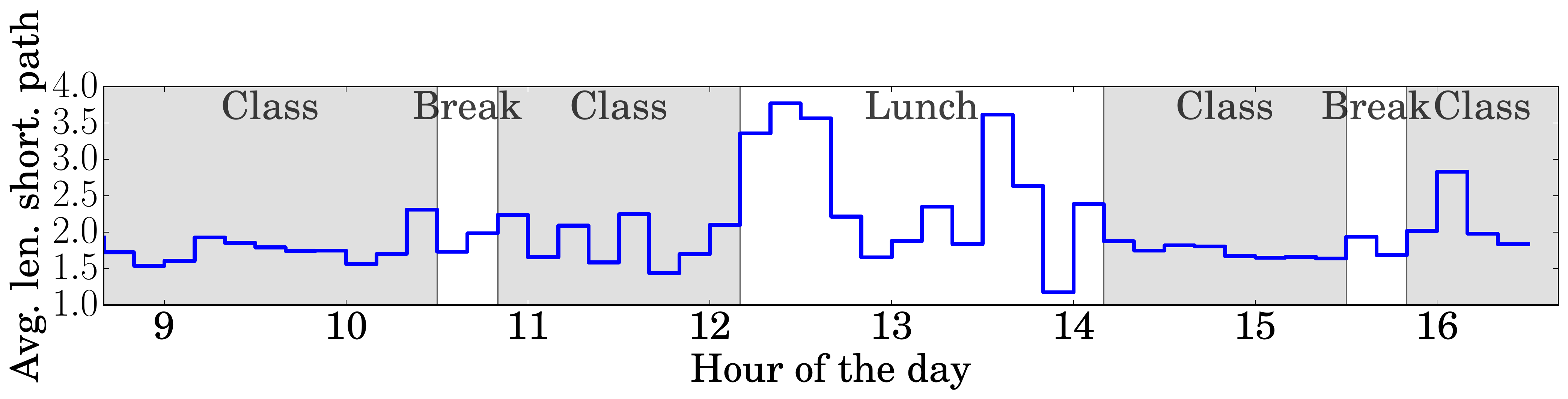}
	}
	
	\caption{\label{fig:prim600_descriptors}Descriptors of the Primary School 
Temporal Network over time. The shaded regions correspond to class periods, 
while white regions correspond to breaks and lunch, according to the information 
given in \cite{Stehle2011}. No significant information are provided by the usual 
network-based descriptors}
\end{figure}

Figure~\ref{fig:prim600_descriptors} shows the network-based descriptors used to 
characterize the temporal network. 
Figures~\ref{subfig:prim600_descriptors_nedges} and 
\ref{subfig:prim600_descriptors_isol} show that the number of edges in the 
temporal network, as well as the number of isolated vertices, is not constant 
over time, reflecting the real-world nature of data: during the lunch break, 
some children leave the school to have lunch at home. The network-based 
descriptors (Figures~\ref{subfig:prim600_descriptors_clust} and 
Figures~\ref{subfig:prim600_descriptors_lensp}) do not provide significant 
insights about the structure of the temporal network.

\subsection{Spectral analysis}

\begin{figure}

	\centering
	\subfloat[\label{subfig:prim600_marginals_freq}Averaged over frequencies 
$\bm{E}_f(c)$ for $c \in 
\{0, 1, 2, 70\}$]{	
		\includegraphics[width=0.95\columnwidth]{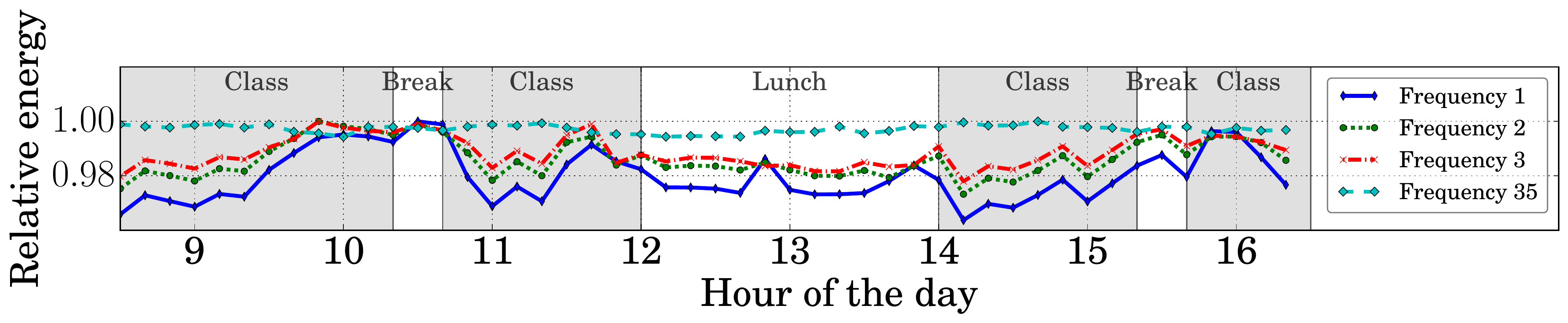}
	}
	
	\subfloat[\label{subfig:prim600_marginals_comp}Averaged over components 
$\bm{E}_c(f)$ for $f \in 
\{1, 2, 3, 35\}$]{	
		\includegraphics[width=0.95\columnwidth]{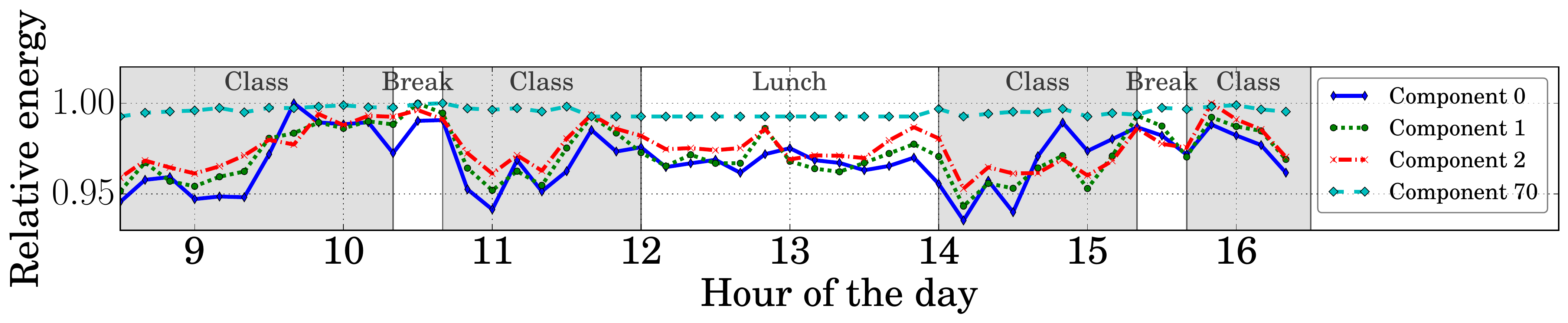}
	}
	
	\caption{\label{fig:prim600_marginals}Marginals of the energies of temporal 
spectra.The energies of the low frequencies and of the first components are not 
equally distributed over time, indicating changes in the global structure of 
the temporal network.}

\end{figure}

Figure~\ref{fig:prim600_marginals} shows the marginals of the energies of 
temporal spectra obtained from the transformation of the primary school temporal 
network. The energies of the low frequencies and of the first components are not 
equally distributed over time: this indicates changes in the global structure of 
the temporal network, that occurs in break periods as well as during lunch. We 
can then divide the day into two main periods: the period where the children 
have class (shaded regions) and the breaks and lunch (white regions).

\begin{figure}

	\centering
	\includegraphics[width=0.95\columnwidth]{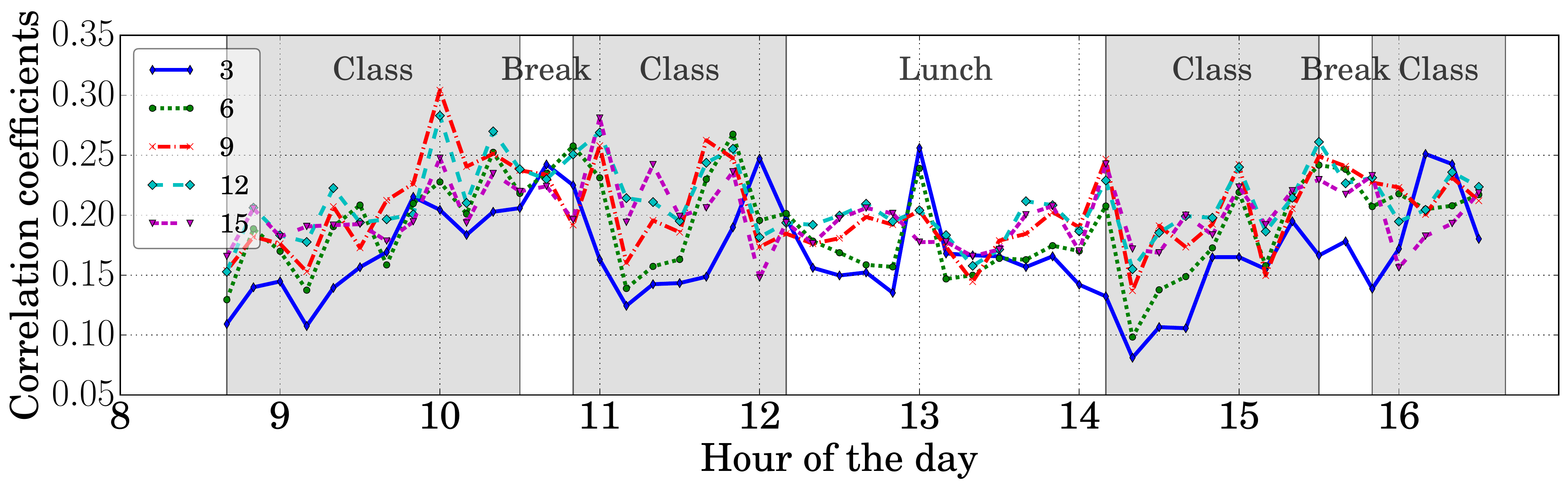}

	\caption{\label{fig:prim600_corr}Correlation between the temporal spectra at 
each time step and the spectra of a network with a network with communities. 
Each line represent the number of communities, averaged over $20$ repetitions. 
The temporal network is correlated with structure with a large number of 
communities during class periods, and with structure with a small number of 
communities during breaks and lunch.}

\end{figure}

Comparing the Primary School Temporal Network with a network with communities 
using correlations between spectra shows that the former temporal network is 
correlated with the latter network involving a large number of communities 
(between 9 and 15) during class periods, a smaller one (between 3 and 6) during 
breaks and lunch periods (see Figure~\ref{fig:prim600_corr}). This is consistent 
with the spatiotemporal trajectories given in \cite{Stehle2011}, 
showing the location of the classes over time: during class periods, the classes 
are separated into different classrooms, while during breaks and lunch, the 
classes mix, yet in two distinct groups.

\subsection{Decomposition into sub-networks}

In the light of the above study of the temporal spectra, we can go further and 
decompose the Primary School Temporal Network into two temporal sub-networks. 
Furthermore, this allows for a quicker interpretation of the obtained 
components, as well as for an evidence of its ability the demonstration of the 
to extract the significant components.

\begin{figure}

	\centering
	\subfloat[\label{subfig:prim600_comp_w}Frequency patterns, obtained after 
reshaping the columns of $\bm{W}$ into matrices.]{
		\includegraphics[width=0.95\columnwidth]{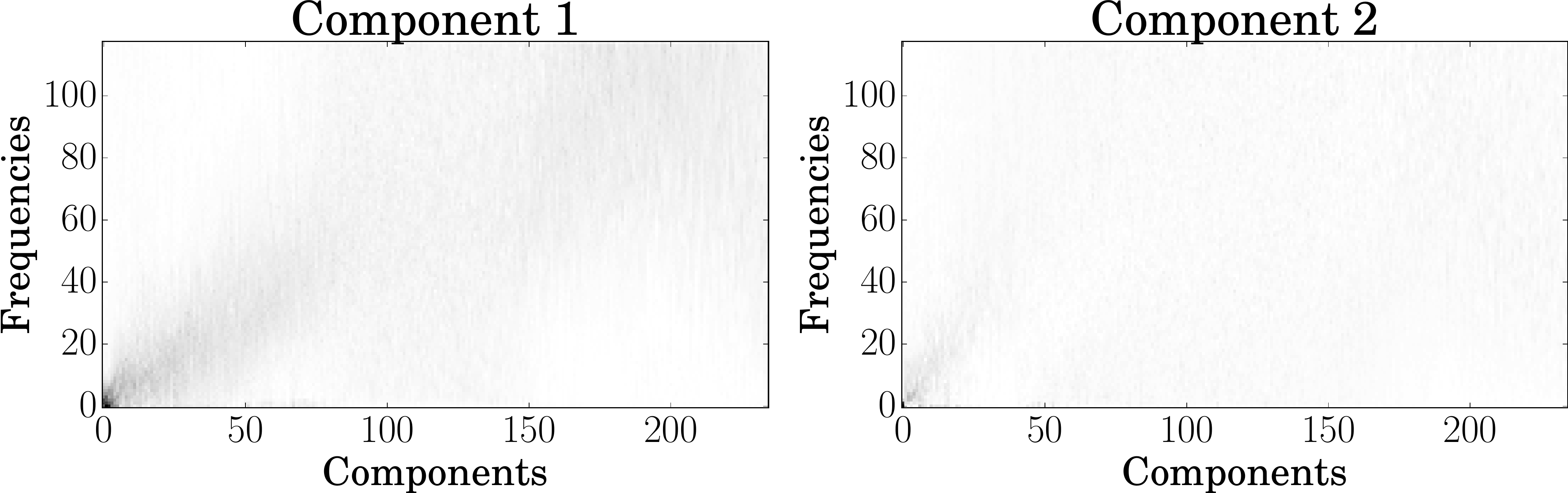}
	}
	
	\subfloat[\label{subfig:prim600_comp_h}Activation coefficients, 
corresponding 
to the rows of the matrix $\bm{H}$, normalized by the maximal value of 
$\bm{H}$]{
		\includegraphics[width=0.95\columnwidth]{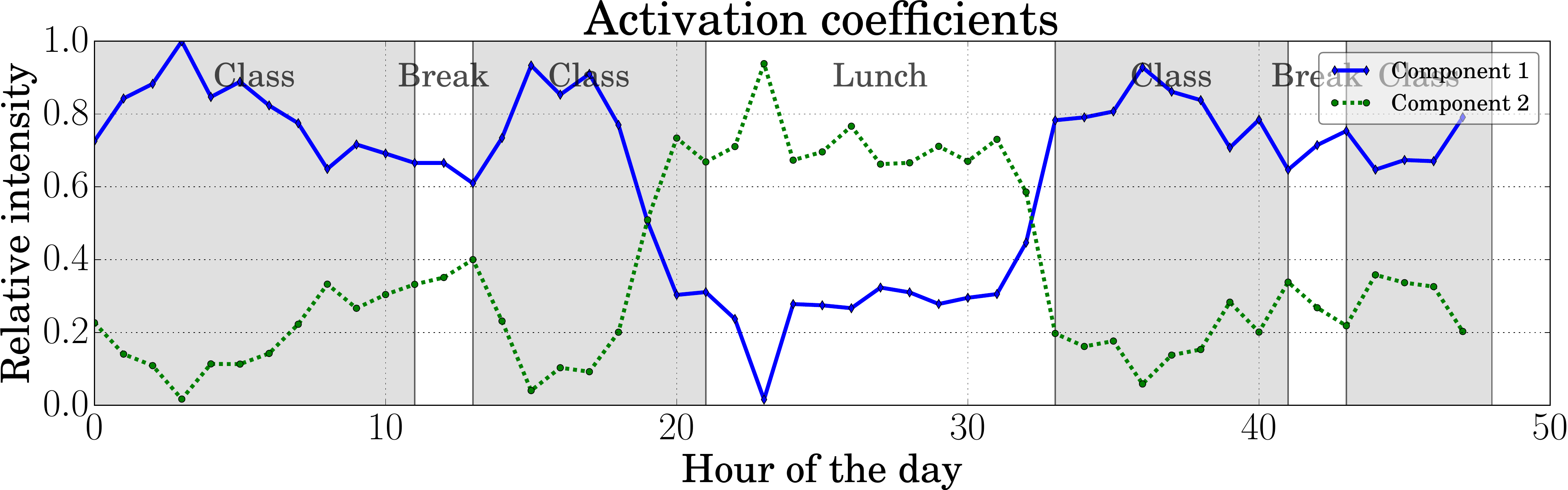}
	}
	\caption{\label{fig:prim600_comp_wh}Results of the nonnegative matrix 
factorization for the Primary School Temporal Network, using $K=2$ and 
$\gamma=5$. The activation coefficients are consistent with the schedule of the 
children in the primary school, as detailed in \cite{Stehle2011}.}

\end{figure}

Figure~\ref{fig:prim600_comp_wh} shows the results of the NMF on the Primary 
School Temporal Network, using $K=2$ and $\gamma=5$. The school day is divided 
into three specific periods, according to the activation coefficients 
(Figure~\ref{subfig:prim600_comp_h}). The first period occurs during class 
hours, where only the component $1$ is mainly active. The second period groups 
together the breaks, during which the components $1$ and $2$ are significantly 
present. Finally, the period $3$ concerns the lunch break, where the component 
$2$ is dominant. This cutting is consistent with the ground truth, as described 
in \cite{Stehle2011}: only two or three classes have breaks at the same time, 
while the other ones stay in their respective classrooms. As for lunches, they 
are taken in two consecutive turns, preserving the structure in classes, with 
nevertheless a weaker intensity.

\begin{figure}

	\centering
	
	\subfloat[\label{subfig:prim600_agg_orig}Original temporal network]{
		\includegraphics[width=0.31\columnwidth]{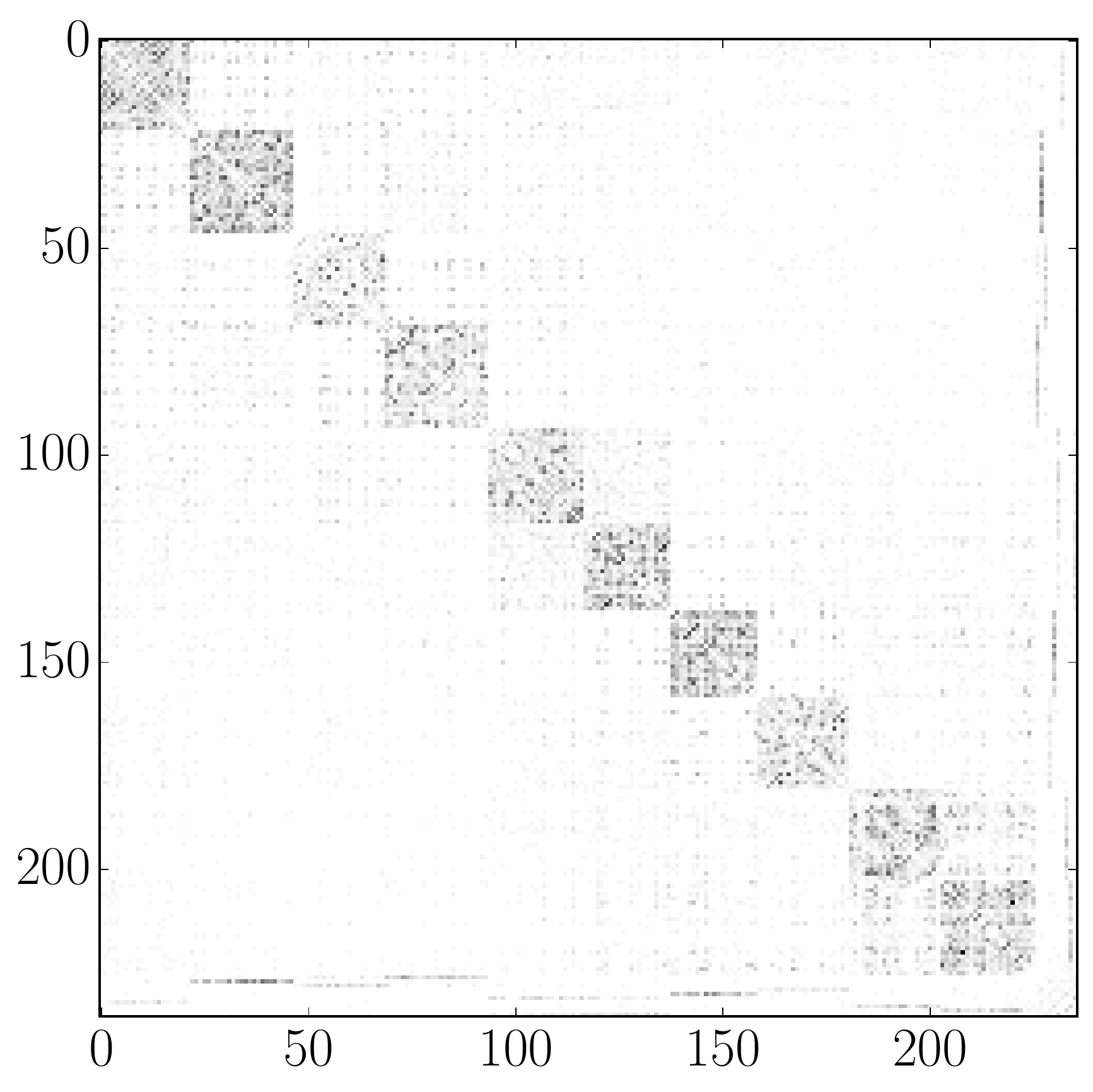}
		\includegraphics[width=0.31\columnwidth]{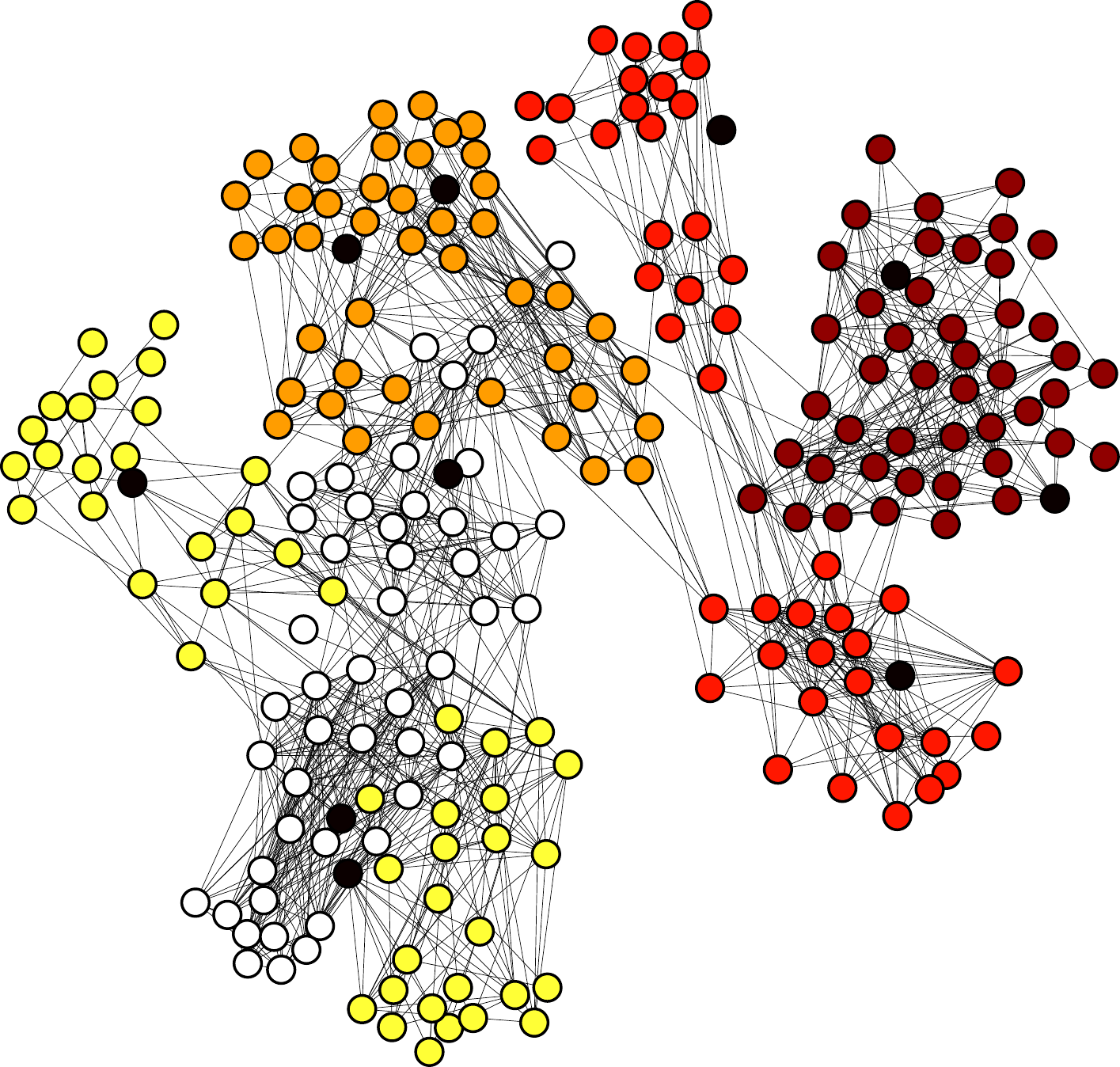}
	\includegraphics[width=0.31\columnwidth]{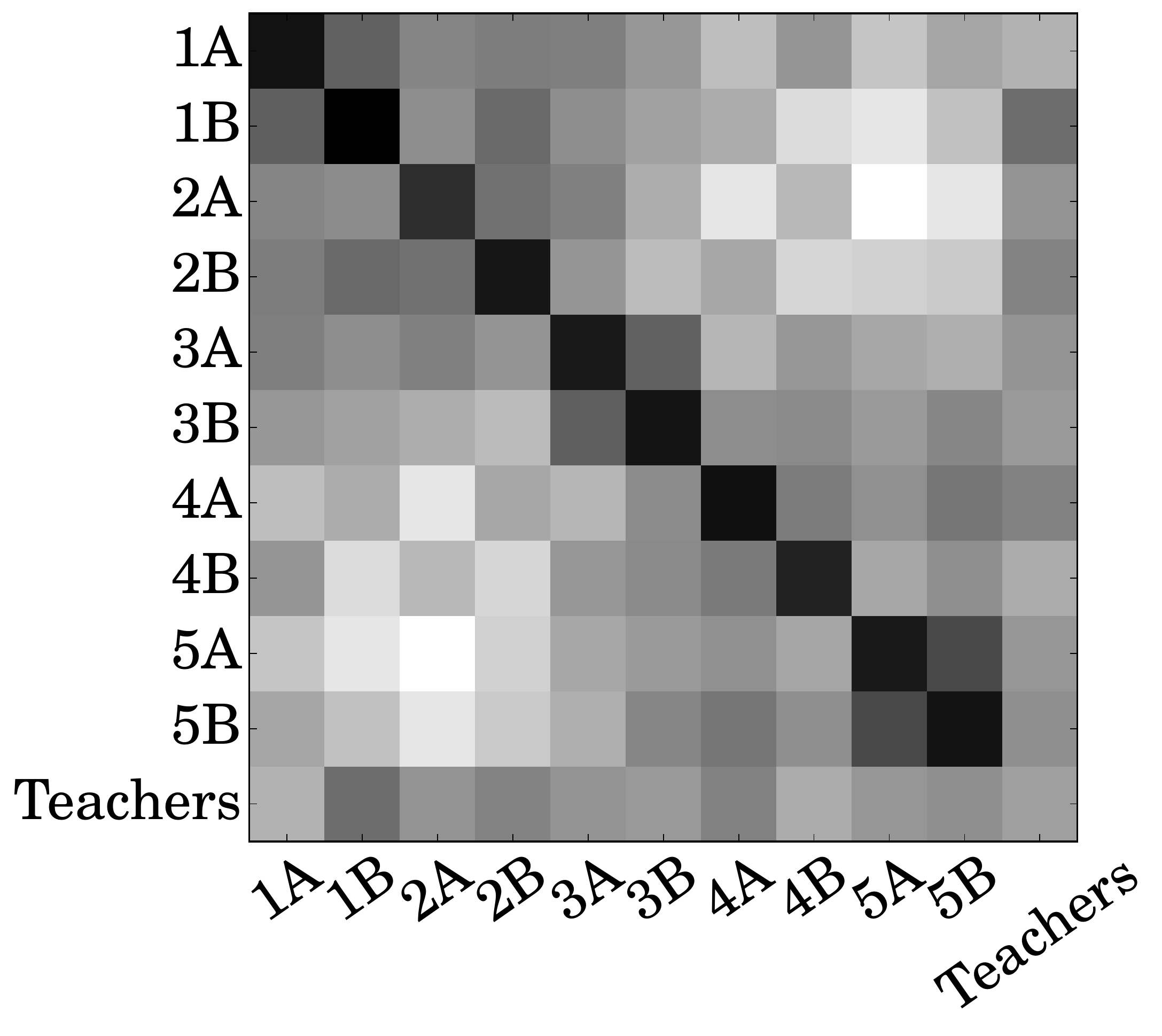}
	}

	\subfloat[\label{subfig:prim600_agg_comp1}Component 1]{
		\includegraphics[width=0.31\columnwidth]{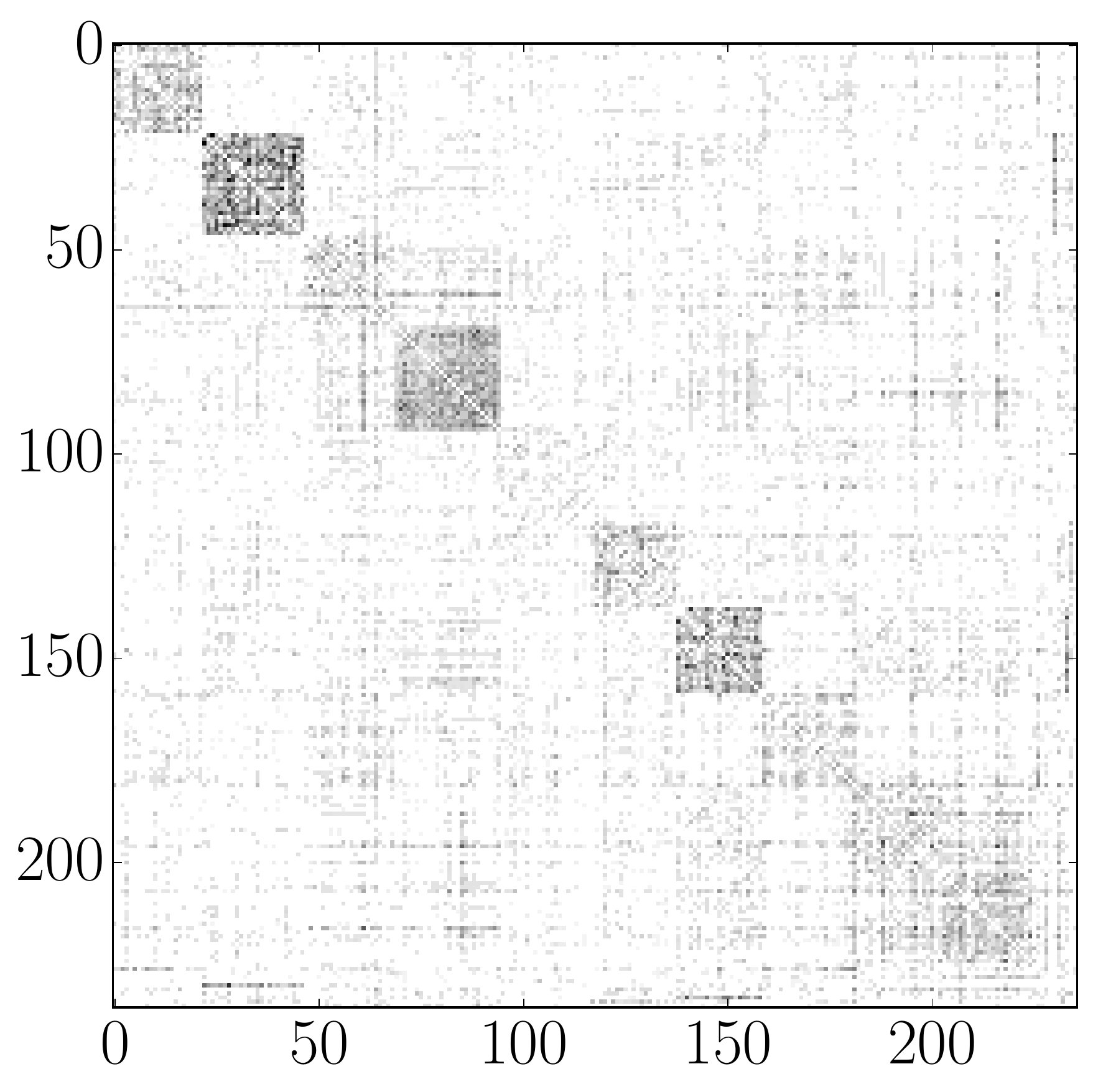}
		\includegraphics[width=0.31\columnwidth]{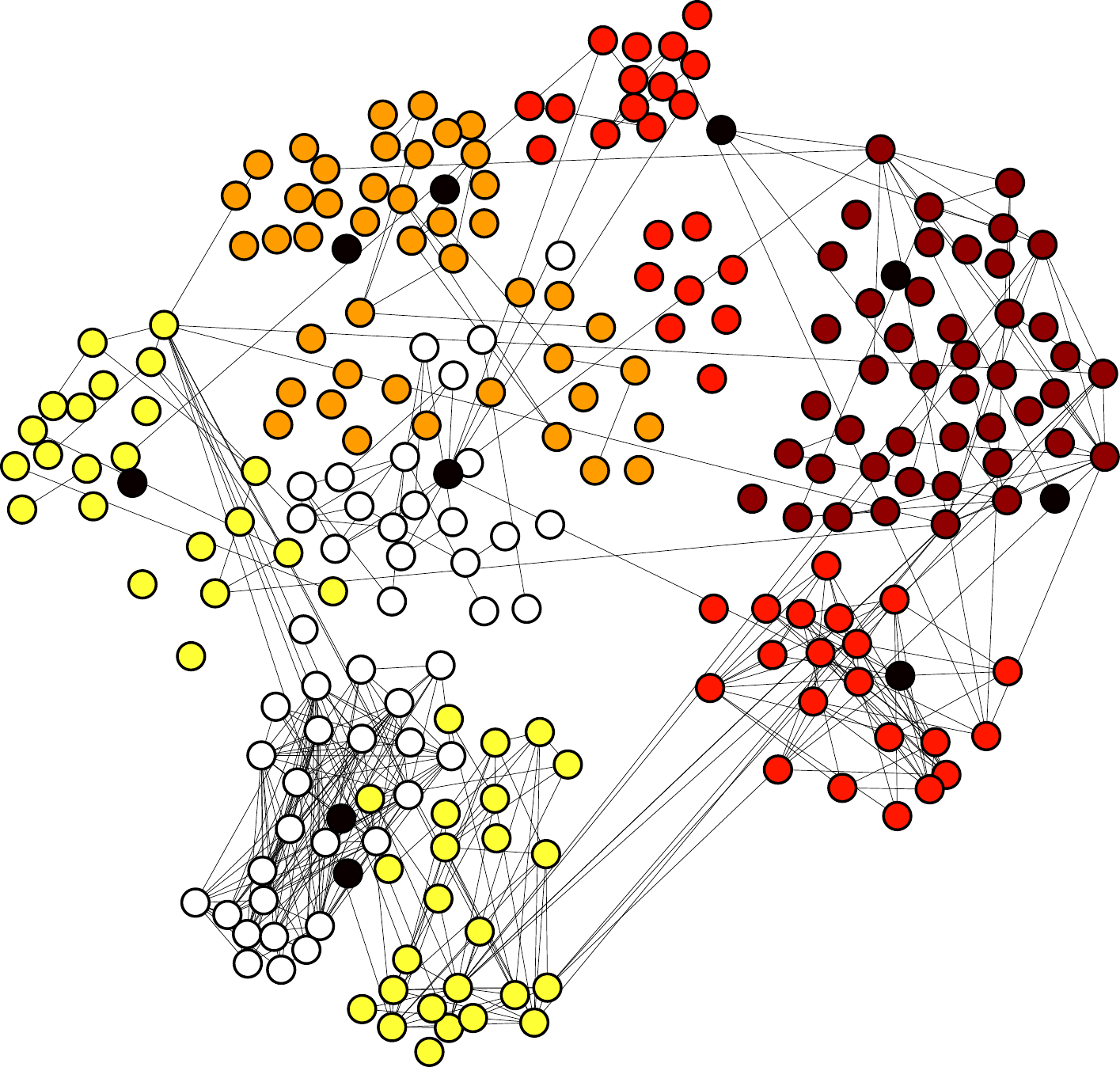}
		\includegraphics[width=0.31\columnwidth]{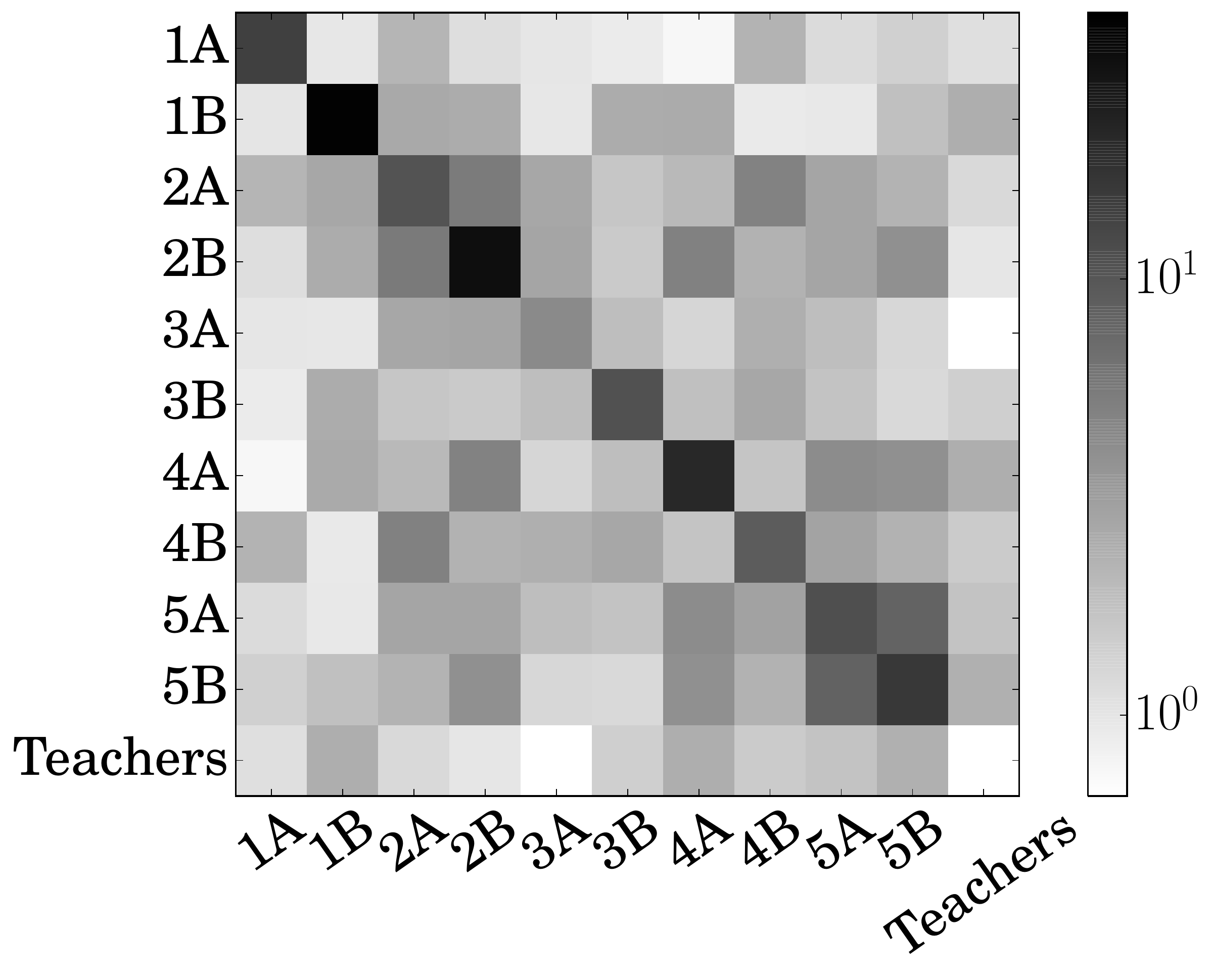}
	}
	
	\subfloat[\label{subfig:prim600_agg_comp2}Component 2]{
		\includegraphics[width=0.31\columnwidth]{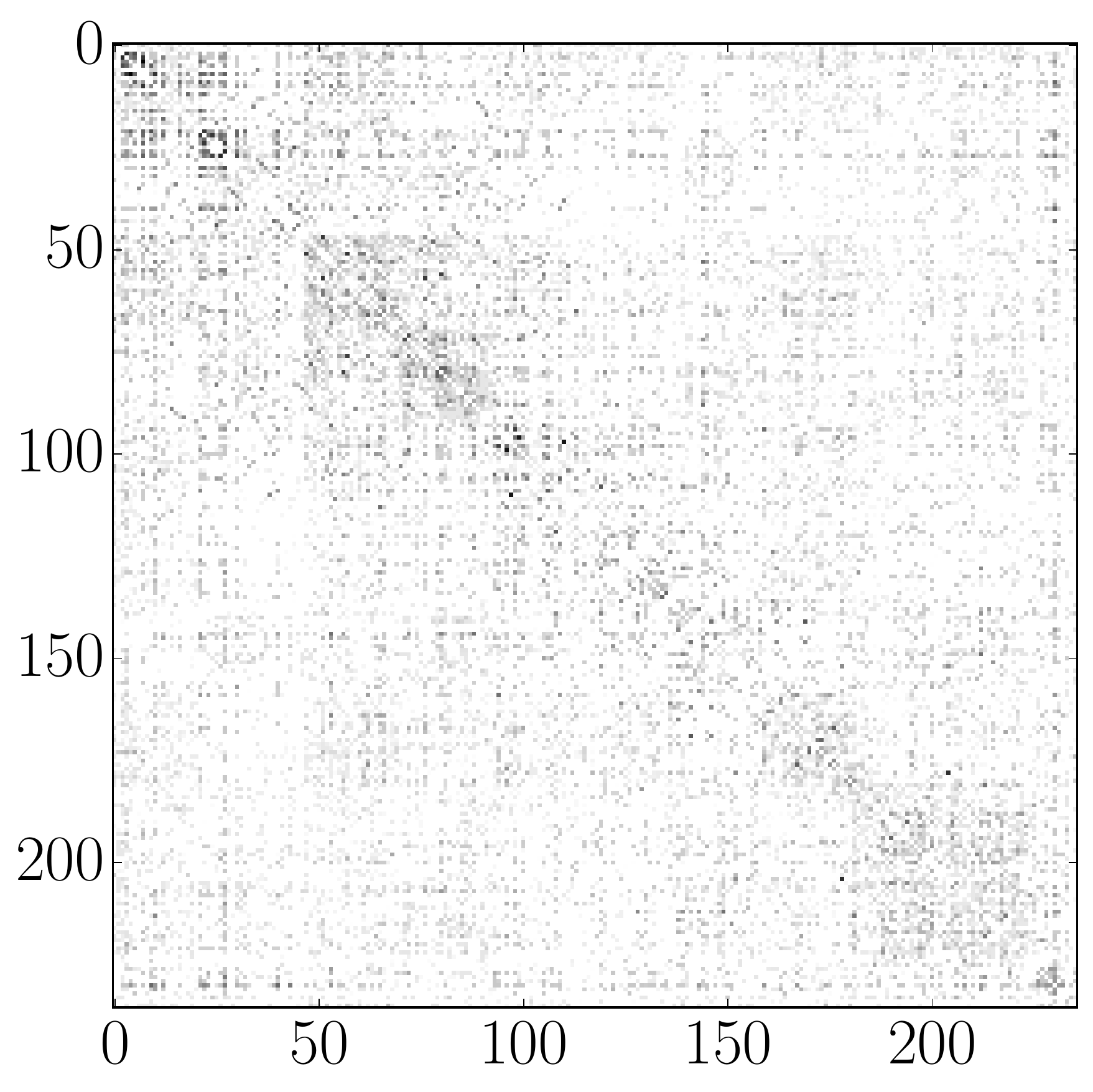}
		\includegraphics[width=0.31\columnwidth]{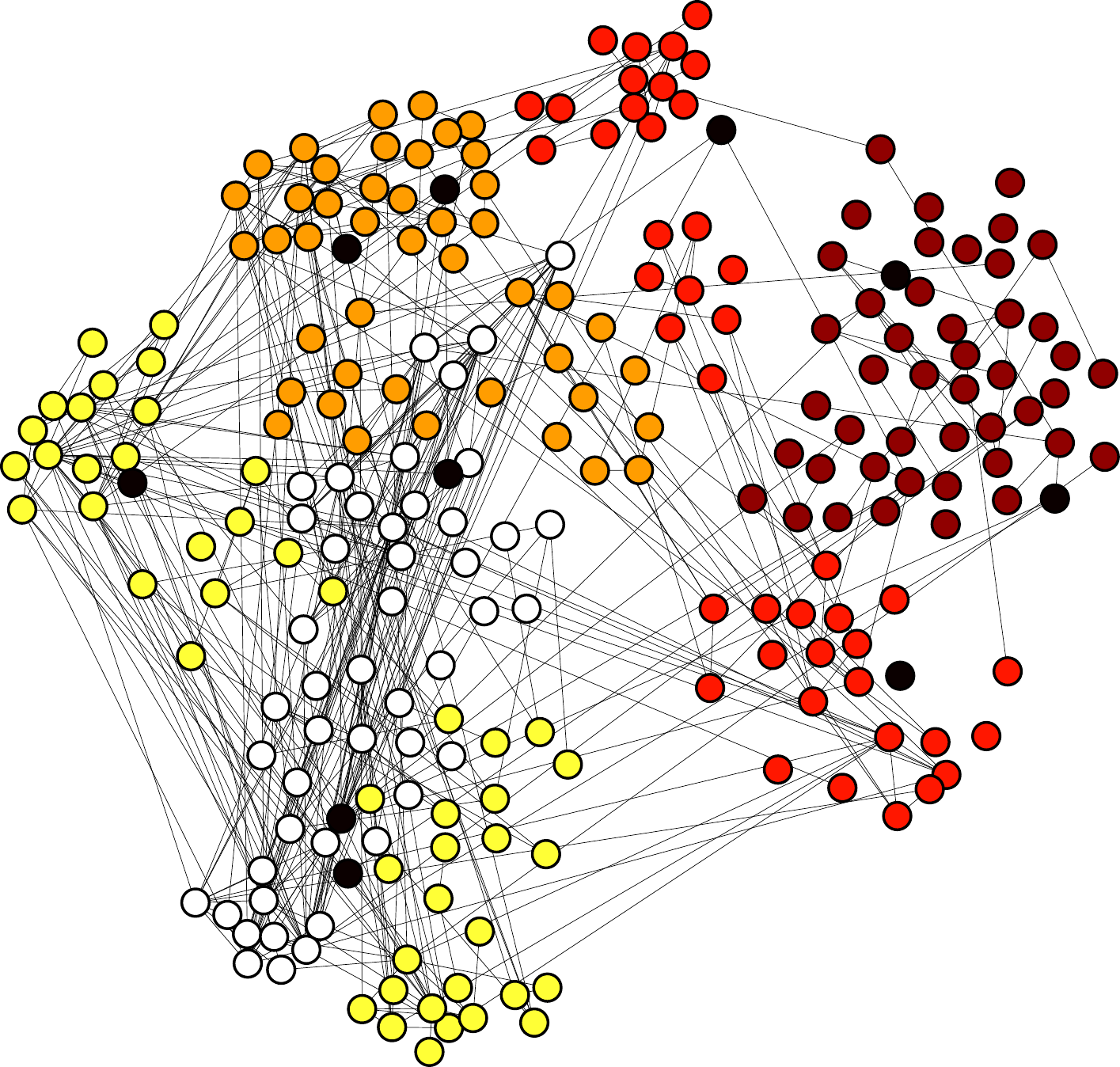}
		\includegraphics[width=0.31\columnwidth]{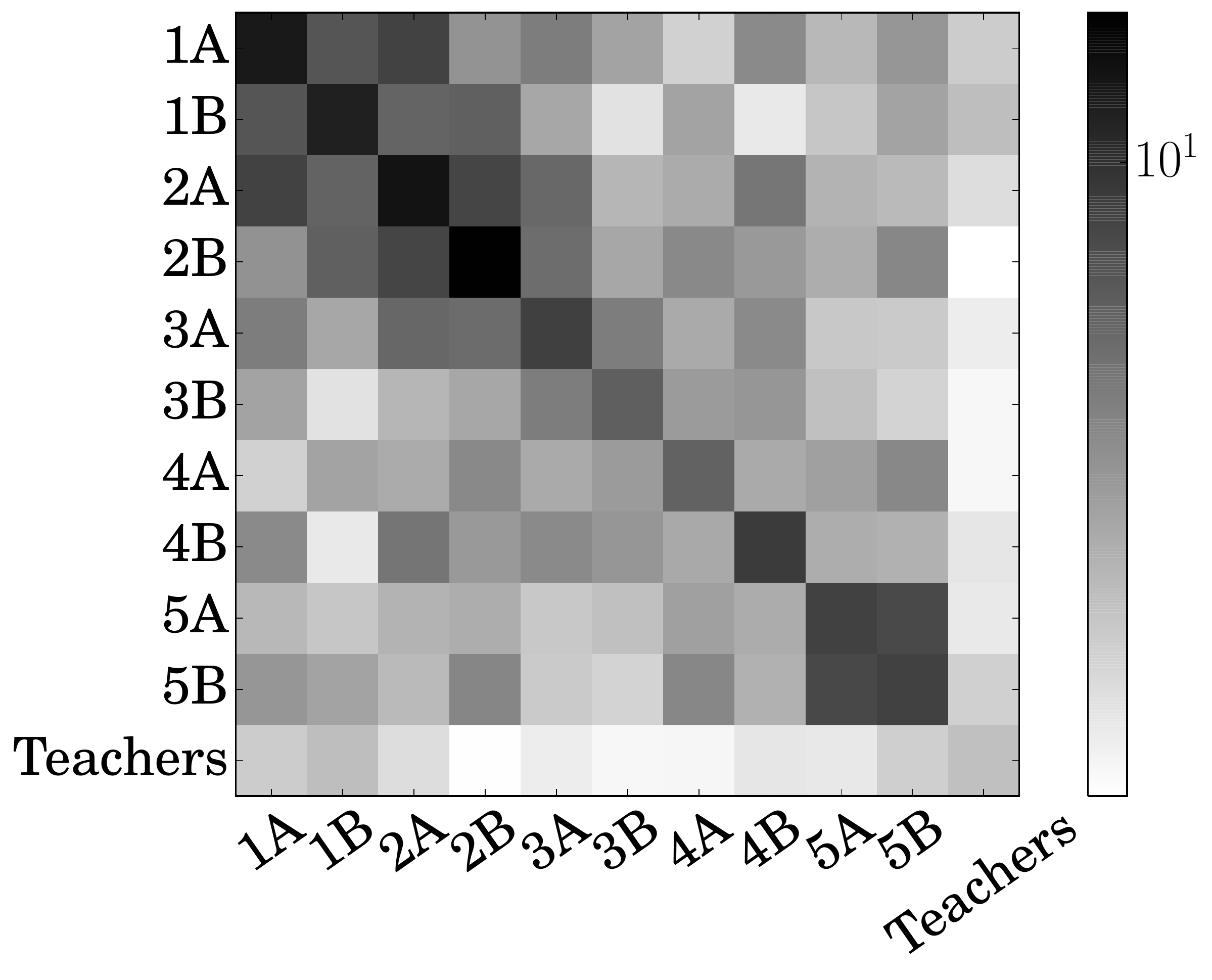}
	}

	\caption{\label{fig:prim600_agg} (Left) Aggregated adjacency matrix over 
time, weighted by the coefficients of $\bm{H}$. (Middle) Network representation 
using the layout provided in \cite{Stehle2011}, after thresholding of edges 
according to their weights. The color of dots indicates the grade of the 
children, while black dots represent the teachers.  (Right) Grayscale-coded 
contact matrix between classes: each entry gives the number of contacts inside 
and between the classes. A logarithmic scale is used to enhance the 
visualization. The component $1$ represents the structures in classes, while the 
component $2$ describes the structure during the breaks and lunch.}

\end{figure}

Figure~\ref{fig:prim600_agg} shows different representations of the temporal 
networks reconstructed from the components. The left figure shows the aggregated 
adjacency matrix over time, as described in Section~\ref{subsec:nmf_ttn}. The 
vertices are ordered according to the classes of the children, from the youngest 
to the oldest. The middle figure shows the aggregated network, using the layout 
provided in \cite{Stehle2011}, after thresholding of edges according to their 
weights. The color of dots indicates the grade of the children, while black dots 
represent the teachers. Finally, the right figure shows the contact matrix 
between classes, obtained by counting the number of edges inside and between the 
classes. A logarithmic scale is used to enhance the visualization. 
Figure~\ref{subfig:prim600_agg_orig} shows the original temporal network, while 
Figures~\ref{subfig:prim600_agg_comp1} and \ref{subfig:prim600_agg_comp2} show 
respectively the component $1$ and the component $2$. We can easily observe that 
the component $1$ describes the structures in classes, with higher density of 
edges inside classes than between classes. Conversely, the component $2$ 
highlights a less structured network pattern, which looks like two communities, 
separating the youngest classes from the oldest. Those observations are 
consistent with the description of lunches mentioned above.

These results highlight the interest of decomposing a temporal network into 
several sub-networks, which can be studied independently of one another.  
Without prior knowledge, the different periods of activity in the primary school 
are displayed, and can then guide the analysis of the system  by restricting the 
analysis over several time intervals.


\section{Conclusion}

We have proposed a novel method to track the structure of temporal networks over 
time using the duality between graphs and signals, as well as classical signal 
processing techniques, such as spectral analysis and nonnegative matrix 
factorization. At each time, the temporal network is represented by a graph 
which is transformed into a collection of signals. NMF is used to extract 
patterns in the energies of spectra of these signals; these patterns, whose 
activation coefficients vary over time, represent a specific structure of the 
underlying network. The effectiveness of the method has been demonstrated on a 
toy example, containing three types of structures as well as on a real-world 
network describing temporal contacts between children in a primary school. 

These results provide insights in the characterization of temporal networks, but 
also call for further studies: In particular, it would be interesting to have a 
deeper look on how the reconstructed temporal networks are embedded in the 
original temporal network. Furthermore, it would be worth considering an 
iterative analysis, by studying the reconstructed temporal network themselves, 
using the same process. This could lead to a spatiotemporal multiscale analysis, 
permitting an increasingly precise track of the structure of the temporal 
network.

\section*{Acknowledgment}
This work is supported by the programs ARC 5 and ARC 6 of the r\'egion 
Rh\^one-Alpes and the project Vél'Innov ANR-12-SOIN-0001-02. Cédric Févotte is 
gratefully acknowledged for discussions and comments.

\bibliographystyle{plain}
\bibliography{bibliography}

\end{document}